\documentclass[pra,aps,amsmath,amssymb,amsfonts,twocolumn,nofootinbib,longbibliography]{revtex4-1}
\usepackage{}
\usepackage{amssymb}
\usepackage{bm,mathrsfs}
\usepackage{graphicx}
\usepackage{epsfig}
\usepackage{amsmath,bbm}
\usepackage{amsfonts,amssymb}
\usepackage{times}
\usepackage{verbatim}
\usepackage[sort&compress]{natbib}
\usepackage{amsmath}
\usepackage{bm}
\usepackage{float}
\allowdisplaybreaks[4]
\usepackage[colorlinks,breaklinks,linkcolor=blue,anchorcolor=blue,citecolor=blue,urlcolor=magenta]{hyperref}

\usepackage{color}
\definecolor{MyDarkBlue}{rgb}{0,0.08,0.45}
\definecolor{yellow}{rgb}{0.99,0.99,0.70}
\definecolor{white}{rgb}{1.0,1.0,1.0}
\definecolor{black}{rgb}{0.00,0.00,0.00}
\definecolor{green}{rgb}{0.8,0.98,0.83}

\begin{document}
\title{Dressed bound states and non-Markovian dynamics with a whispering-gallery-mode microcavity coupled to a two-level atom and a semi-infinite photonic waveguide}
\author{J. Y. Sun$^{1}$}

\author{C. Cui$^{1}$}

\author{Y. F. Li$^{1}$}

\author{Shuang Xu$^{2}$}

\author{Cheng Shang$^{3,4}$}
\email{Contact author: cheng.shang@riken.jp}

\author{Yan-Hui Zhou$^{5}$}
\email{Contact author: yanhuizhou@126.com}

\author{H. Z. Shen$^{1}$}
\email{Contact author: shenhz458@nenu.edu.cn}

\makeatletter
\renewcommand\frontmatter@affiliationfont{\vspace{1mm} \small}
\makeatother

\affiliation{\textit{$^{1}$Center for Quantum Sciences and School of Physics, Northeast Normal University, Changchun 130024, China\\
$^{2}$College of Sciences, Northeastern University, Shenyang 110819, China\\
$^{3}$Analytical quantum complexity RIKEN Hakubi Research Team, RIKEN Center for Quantum Computing (RQC), Wako, Saitama 351-0198, Japan\\
$^{4}$Department of Physics, The University of Tokyo, 5-1-5 Kashiwanoha, Kashiwa, Chiba 277-8574, Japan\\
$^{5}$Quantum Information Research Center and Jiangxi Province Key Laboratory of Applied Optical Technology, Shangrao Normal University, Shangrao 334001, China}}
\date{\today}

\begin{abstract}
We investigate the dressed bound states in an open cavity with a whispering-gallery-mode (WGM) microring coupled to a two-level atom and a waveguide with a mirror at the right end. We demonstrate that the non-Hermiticity of an open cavity facilitates the formation of the dressed bound states, which consists of the vacancy-like dressed bound states and Friedrich-Wintgen dressed bound states. By deriving analytical conditions for these dressed bound states, we show that when a two-level atom couples to the standing-wave mode that corresponds to a node of the photonic wave function the vacancy-like dressed bound states occur, which are characterized by null spectral density at cavity resonance. Conversely, Friedrich-Wintgen dressed bound states can be realized by continuously adjusting system parameters and indicated by the disappearance of the Rabi peak in the emission spectrum, which is a distinctive feature in the strong-coupling regime. Moreover, we extend our analysis to the non-Markovian regime and find that our results are consistent with those obtained under the Markovian approximation in the wideband limit. In the non-Markovian regime, we analyze dressed bound states for both zero and non-zero accumulated phase factors. For zero accumulated phase factors, the non-Markovian regime exhibits higher peak values and longer relaxation times for vacancy-like dressed bound states compared to the Markovian regime, where the Friedrich-Wintgen dressed bound states are absent in the non-Markovian case. Finally, we establish the correspondence between the energy spectrum and bound state conditions for non-zero accumulated phase factors and analyze the influence of various parameters on non-Markovian bound states. Our work exhibits bound state manipulations through non-Markovian open quantum system, which holds great potential for building high-performance quantum devices for applications such as sensing, photon storage, and nonclassical light generation.
\end{abstract}

%\pacs{42.50.Pq, 42.50.Ct, 42.50.Ar, 42.50.Dv} \maketitle
%42.50.pq: Cavity quantum electrodynamics; micromasers
%42.50.Ct: Quantum description of interaction of light and matter; related experiments
%42.50.Ar: Photon statistics and coherence theory
%42.50.Dv: Quantum state engineering and measurements

\maketitle
\section{Introduction}
%PRL 126, 063601
Atom-photon dressed states are a fundamental concept in quantum electrodynamics (QED) \cite{Cohen1992,Haroche2006}. In particular, a dressed bound state (DBS) features a photonic cloud that remains localized close to the atom. Dramatic departures from spontaneous decay such as vacuum Rabi oscillations \cite{Haroche2006,Raimond565} or population trapping \cite{Bykov861,John1764, Kofman353,Lambropoulos455} thus occur.
%PRA 98, 053440
Theoretical investigations into dressed states began in the late 1970s \cite{Cohen345}, focusing on the resonance fluorescence and absorption spectra of multilevel atoms. The approach determines the quasi-eigenenergy levels of the system of coupled light and electrons as eigenvalues of the Floquet Hamiltonian \cite{Shirley979}. Since then, the dressed-state approach has been applied in analyzing a wide variety of novel physical phenomena in laser physics and chemistry or in quantum optics, such as dynamical Stark effect \cite{Xu227401,Gall057401,Yang12613}, electromagnetically induced transparency \cite{Wu173602,Ian063823,Safavi69}, cavity QED \cite{Blais062320,Pirkkalainen211,Gely245115}, and so on. Moreover, recent advances in strong- and ultrashort-pulsed laser light technology have opened new sophisticated ways of manipulating atoms and molecules relying on dressed states, demonstrated such as in softening or hardening of chemical bonds in photochemical reactions \cite{Bucksbaum1883,Giusti3869,Yao1993,Frasinski3625} and in laser-assisted elastic electron scattering \cite{Morimoto123201}.

%PRA 102,023702
In conventional waveguide QED, where the rotating wave approximation is applicable, the primary aim is to manipulate atom-atom interactions facilitated by the electromagnetic fluctuations within the waveguide \cite{Dzsotjan075419,Gonzalez020501,Zueco024503,Zheng113601,Manzoni14696}. These propagating photons induce interactions that are long-range yet dissipative, with dissipation arising from the loss of information in the traveling wave packets. Conversely, dressed atom-field eigenstates are localized around the quantum emitter and referred to as bound states \cite{John2169,John2486,John2418,John12772,John1764}, which generate nondissipative but exponentially bounded interactions \cite{Bello0297,Sanchez023003,Gonzalez320,Douglas326,Shi021027,Calajo033833,Gonzalez043811,Gonzalez143602,Gonzalez97,Gonzalez043831,Gonzalez221,Shi105005}. These exact nonpropagating eigenstates lie within the band gap, rendering them nonpropagating. Besides, bound states modify the spontaneous emission \cite{Khalfin1053,Bykov861,Fonda587,Onley432,Gaveau7359,Garmon115318,Garmon261,Garmon2013,Lombardo053826,Sanchez023831}, which makes them an interesting resource for engineering quantum photonics.

%yangjinfeng
With the rapid development of quantum information technology \cite{Nielsen2000,Ladd45}, open quantum systems \cite{Breuer2002,Weiss2008,Shen052122} have attracted increasing attention. In general, all quantum systems in reality are open owing to the unavoidable coupling with the environments \cite{Caruso1203,Gardiner2000,Li062124,Franco1345053}. The Markovian approximation for open systems \cite{Breuer2002,Weiss2008} is only valid when the coupling between the system and environment is weak and the characteristic times of the system under study are significantly larger than those of the bath. Otherwise we should take the non-Markovian effects generated by the environment acting on the system dynamics \cite{Reuther062123,Cui032209,Shen033805} into account, which occur in many quantum systems including the coupled cavities \cite{Link020348,Yang053712}, atom-cavity systems \cite{Shen053705,Shen032101,Shen033835}, photonic waveguides \cite{Li023712,Xin053706}, optomechanical systems \cite{Zhang033701}, periodically driven systems \cite{Shen2852}, photonic crystals \cite{Burgess062207,Hoeppe043603,Shen012107}, colored noises \cite{Costa052126}, cavities coupled to waveguides \cite{Chang052105,Longhi063826,Tan032102,Vega015001}, and implemented in experiments \cite{Liu931,Xiong032101,Cialdi052104,Khurana022107, Madsen233601,Guo230401,Li140501,Xu042328,Tang10002,Uriri052107,Anderson3202,Liu062208}. The non-Markovian process are proving to be valuable in quantum information processing applications, including quantum channel capacity \cite{Bylicka5720,Xue052304}, photon blockade \cite{Shen043714,Shen013826,Shen023856}, quantum batteries \cite{Li5614}, dispersive readout \cite{Shen023707}, bound states \cite{Shen31504,Cui042129} and quantum Brownian motion \cite{Shen042121}. The non-Markovian effects of the environments back-acting on the system dynamics can be characterized by the excitation backflowing between the system and its environment \cite{Breuer021002,Breuer210401,Laine062115,Addis052103,Wibmann062108,Wibmann042108}, which leads to different measures of non-Markovianity \cite{Lorenzo020102,Rivas050403,Luo044101,Wolf150402,Lu042103,Chruscinski120404}. The insight mentioned above inspire us to investigate the creation and control of dressed bound states (DBSs) within an open cavity affected by non-Markovian effects. 

In this paper, we investigate the DBSs of the optical cavity with a whispering-gallery-mode (WGM) microring coupled to a two-level atom and a waveguide with a mirror at the right end. Under the Markovian approximation, we derive the analytical conditions for the existence of DBSs, whose origin is revealed. Our results show that DBSs in the optical cavity can be categorized into two distinct types: the vacancy-like DBS and the Friedrich-Wintgen DBS. The vacancy-like DBS is characterized by its independence from the atom-photon coupling strength as the cavity mode coupled to the two-level atom coincides with a node in the photonic wave function. In contrast, the Friedrich-Wintgen DBS is contingent on system parameters, such as frequency detuning and coupling strength between different components, which must satisfy the condition for destructive interference between two coupling pathways. The dressed bound states are then generalized to the non-Markovian regimes and compared with Markovian approximation in the wide-band limit.
In the non-Markovian regime, we investigate dressed bound states for both zero and non-zero accumulated phase factors. For zero accumulated phase factors, the peak value and relaxation time of the vacancy-like dressed bound state are higher in the non-Markovian regime compared to the Markovian regime. Additionally, Friedrich-Wintgen dressed bound states are absent in the non-Markovian case. Moreover, we establish the correspondence between the energy spectrum and bound state conditions for non-zero accumulated phase factors and examine the influences of various parameters on non-Markovian bound states.

The remainder of the paper is organized as follows. Sec. II describes the theoretical model and Hamiltonian for the optical cavity with a WGM microring coupled to a two-level atom. In Sec. III, the results of the DBSs under the Markovian approximation are analytically and numerically discussed. Subsequently, in Sec. IV, we extend the DBSs in the optical cavity to a non-Markovian bath in the case of zero and non-zero accumulated phase factor. Finally, a summary of the main results is given in Sec. V.

\section{Hamiltonian and dynamics with the Markovian approximation}

\begin{figure}[tbp]
   \centerline{
   \includegraphics[width=7.8cm, height=6cm, clip]{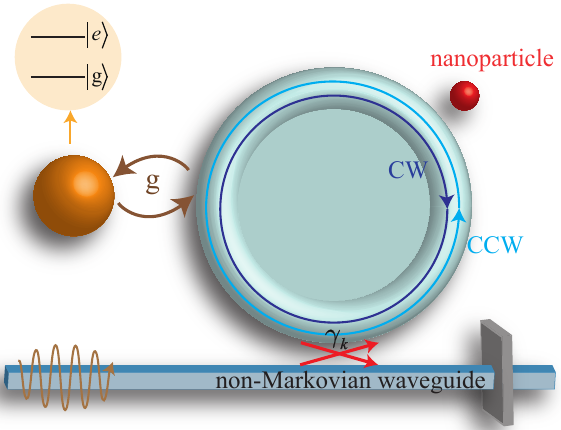}}
   \caption{Schematic of the optical cavity with a WGM with clockwise (CW) and counterclockwise (CCW) propagating modes microring coupled to a two-level atom (excited state $| e \rangle $ and ground state $| g \rangle $) and a non-Markovian waveguide via a mirror at the right end, where $\gamma_k$ denotes the coupling coefficiant between the cavity and waveguide. The waveguide is driven by a laser with frequency $\omega$ to the WGM microresonator through a tapered fiber. The taper-scattering nanoparticle induced dissipative back-scattering leads to the coupling between the CW and CCW modes. The parameter $g$ denotes the coupling strength between the WGM optical cavity and the linearly polarized atom.}\label{device}
 \end{figure}

%PRA 107, 053705
As schematically shown in Fig.~\ref{device}, the model we consider is a WGM microdisk resonator with a single nanoparticle (or scatterer) within its mode volume. The particle in the evanescent field of the resonator acts as a scatterer and induces the coupling between the clockwise (CW) and counterclockwise (CCW) propagating modes with strength $J$, resulting in optical mode splitting \cite{Zhu46}.
%jinghui guan yu particle de tuo zhan
%Mode splitting induced by other factors such as surface roughness  or material inhomogeneity can be pre-detected and thus minimized [8,10,26], or modes and resonators without such intrinsic  mode splitting can be used.  We note that by directly revealing  particle polarizability (a parameter depending on the size, shape,  and refractive index of the particle) [8], this technique can discriminate between particles of the same size but different refractive indices or shapes [8,10].
This resonator, characterized by optical resonance frequency $\omega_c$ and intrinsic loss $\gamma$, is coupled to a semi-infinite waveguide with a perfect mirror (i.e., unity reflectivity) at the right end. A linearly polarized two-level atom with frequency $\omega_0$ couples to the WGM optical cavity with coupling strength $g$. We assume that the atom is embedded within the cavity, thereby suppressing its coupling to free space via modes other than the cavity modes. Furthermore, although the optical cavity supports a series of WGM resonances, we focus exclusively on a single pair of degenerate CW and CCW modes. This simplification is justified since the linewidth of the atom can be much smaller than the frequency spacing $\Delta f$ between the adjacent WGM resonances in a realistic optical cavity. For instance, $\Delta f$ is estimated to be approximately $\sim 11$ THz for a SiN microdisk with a $2$ $\rm{\mu m}$ radius \cite{Doeleman1943,Lu115434}, while the linewidth of CdSe/ZnSe quantum dots is approximately $\sim 1$ THz at $77$ K \cite{Odoi2769, Leistikow045301}.

%figure 1

The Hamiltonian of the total system in the Markovian approximation ($\gamma_k=\sqrt{\frac{\gamma }{2\pi }}$) reads (setting $\hbar  \equiv 1$) \cite{Carmichael2273,Mok013369,Lu043714}
\begin{equation}
\begin{aligned}
  \hat{H} = {\hat{H}_S} + {\hat{H}_B} + {\hat{H}_{SB}},
\label{HamiltonianTotal}
\end{aligned}
\end{equation}
with
\begin{equation}
\begin{aligned}
  {\hat{H}_S} =& {\omega _0}{\hat{\sigma} _ + }{\hat{\sigma} _ - } + {\omega _c}\hat{c}_{\mathrm{cw}}^\dag {\hat{c}_{\mathrm{cw}}} + {\omega _c}\hat{c}_{\mathrm{ccw}}^\dag {\hat{c}_{\mathrm{ccw}}} \\
              &+ J(\hat{c}_{\mathrm{cw}}^\dag {\hat{c}_{\mathrm{ccw}}} + \hat{c}_{\mathrm{ccw}}^\dag {\hat{c}_{\mathrm{cw}}})\\
              &+ \sum\limits_{j = \mathrm{cw},\mathrm{ccw}} {g(\hat{c}_j^\dag {\hat{\sigma} _ - } + {\hat{\sigma} _ + }{\hat{c}_j})},  \\
  {\hat{H}_B} =& \int {d\omega \omega \hat{b}_k^\dag {\hat{b}_k} },\\
  {\hat{H}_{SB}} =& i\sum\limits_{j = \mathrm{cw},\mathrm{ccw}} {\int {d\omega \sqrt{\frac{\gamma }{{2\pi }}} \hat{b}_k^\dag {e^{ - ik{x_j}}}{\hat{c}_j}}  + h.c.},
\label{HamiltonianExpand}
\end{aligned}
\end{equation}
where ${\hat{H}_B}$ describes the free Hamiltonian of the waveguide. ${\hat{H}_{SB}}$ is the Hamiltonian of the cavity-waveguide interaction. ${\hat{\sigma} _ - }$ denotes the lowering operator of the two-level atom. The bosonic annihilation operators for the counterclockwise (CCW) and clockwise (CW) modes are represented by $\hat{c}_{\mathrm{ccw}}$ and $\hat{c}_{\mathrm{cw}}$, respectively. ${\hat{b}_k}$ represents the bosonic annihilation operator of the right-propagating waveguide mode, characterized by frequency $\omega$ and wave vector $k = \omega / v$, where $v$ is the group velocity. $x_{\mathrm{ccw}}$ and $x_{\mathrm{cw}}$ denote the positions of the CCW mode and the mirrored CW mode, respectively.

Hamiltonian~(\ref{HamiltonianTotal}) is analytically solvable due to the total excitation number $\hat{N}={\hat{\sigma}_+}{\hat{\sigma}_-} + \hat{c}_{\mathrm{cw}}^\dag {\hat{c}_{\mathrm{cw}}} + \hat{c}_{\mathrm{ccw}}^\dag {\hat{c}_{\mathrm{ccw}}} + \int {d\omega \hat{b}_k^\dag {\hat{b}_k} }$ being conserved. We propose a scheme that in the spontaneous emission process there is at most one photon in the system under the condition for atom and waveguide respectively initialization in the excited state and the vacuum state 
\begin{equation}
\begin{aligned}
  | {\psi (0)} \rangle  = | {e,0,0,{0}} \rangle,
\label{InitialState}
\end{aligned}
\end{equation}
where $| {e,0,0,{0}} \rangle = | e \rangle \otimes | 0 \rangle_{\mathrm{cw}} \otimes | 0 \rangle_{\mathrm{ccw}} \otimes | 0 \rangle $. $| 0 \rangle$ denotes that all the modes of the environment are in the vacuum state, i.e., $| 0 \rangle = |0\rangle_1 \otimes |0\rangle_2 \ldots \otimes |0\rangle_\infty$. The quantum state of the total system can be expressed as
\begin{equation}
\begin{aligned}
\left| {\psi (t)} \right\rangle  = &A(t)\left| {{{e,0,0,}}{{{0}}}} \right\rangle  + B(t)\left| {{{g,1,0,}}{{{0}}}} \right\rangle  + C(t)\left| {{{g,0,1,}}{{{0}}}} \right\rangle  \\
& + \int {d\omega D_k(t)\hat{b}_k^\dag  \left| {{{g,0,0,}}{{{0}}}} \right\rangle },
\label{expressstate}
\end{aligned}
\end{equation}
where $\hat{b}_k^\dag  \left| {{{g,0,0,}}{{{0}}}} \right\rangle$ signifies that the waveguide contains only one excitation in the $k$th field mode with the corresponding probability amplitude $|D_k(t)|^2$. Substituting Eq.~(\ref{expressstate}) into Schr\"{o}dinger equation $i|\dot{\psi}(t)\rangle=\hat{H}|\psi(t)\rangle$, we obtain a set of the differential equations for the probability amplitudes
\begin{equation}
\begin{aligned}
i\dot A(t) &= ({\omega _0} - {\omega _c})A(t) + gB(t) + gC(t),\\
i\dot B(t) &= gA(t) - i\frac{\gamma }{2}B(t) + (J - i\gamma {e^{i\phi }})C(t),\\
i\dot C(t) &= gA(t) + JB(t) - i\frac{\gamma }{2}C(t),
\label{dotABCt}
\end{aligned}
\end{equation}
where $\phi=\omega_c (x_{cw}-x_{ccw})/v$ denotes the accumulated phase factor of light propagation. It is worth noting that the accumulated phase factor is positive, i.e., $\phi>0$ due to the presence of mirrors on the right side of the waveguide. The derivation details of Eq.~(\ref{dotABCt}) can be found in Appendix \ref{APPA}. Considering the expectation value of arbitrary operator $\hat{A}$ defined by $\langle \hat{A} \rangle  = \left\langle {\psi (t)} \right|\hat{A} \left| {\psi (t)} \right\rangle$, we get the equations of motion in the single excitation subspace
\begin{equation}
\begin{aligned}
\frac{d}{d t} \vec{\varepsilon}=-i \mathbf{Q}_{c} \vec{\varepsilon},
\label{deltap}
\end{aligned}
\end{equation}
 where $\vec{\varepsilon}=\left[\left\langle\hat\sigma_{-}\right\rangle,\left\langle\hat c_{\mathrm{ccw}}\right\rangle,\left\langle\hat c_{\mathrm{cw}}\right\rangle\right]^{T}$ and the matrix
\begin{equation}
\begin{aligned}
\mathbf{Q}_{c}=\left[ {\begin{array}{*{20}{c}}
  {{\omega _0} - {\omega _c}}&g&g \\
  g&{ - i\frac{\gamma}{2}}&{J - i\gamma{e^{i\phi }}} \\
  g&J&{ - i\frac{\gamma}{2}}
\end{array}} \right].
\label{Mc}
\end{aligned}
\end{equation}

We investigate the spectrum properties of DBSs through the spontaneous emission spectrum of the atom, which can be measured via fluorescence of the $\mathrm{atom}$. Here we define the emission spectrum as $S(\omega ) = \mathop {\lim }\limits_{t \to \infty } \operatorname{Re} \left[ {\int_0^\infty  d \tau \left\langle {{{\hat \sigma }_ + }(t){{\hat \sigma }_ - }(t + \tau )} \right\rangle {e^{i\omega \tau }}} \right]$ \cite{Zhu46}, which allows for experimental measurement through atomic fluorescence. It is straightforward to demonstrate
\begin{equation}
\begin{aligned}
\left\langle {{{\hat \sigma }_ + }(0){{\hat \sigma }_ - }(\tau )} \right\rangle {{ }} &= \langle \psi (0)|{{\hat \sigma }_ + }(0){e^{i\hat H\tau }}{{\hat \sigma }_ - }(0){e^{ - i\hat H\tau }}\left| {\psi (0)} \right\rangle  \\
&=\langle {{g,0,0,}}{{{0}}}|{e^{i\hat H\tau }}{{\hat \sigma }_ - }(0){e^{ - i\hat H\tau }}\left| {{{e,0,0,}}{{{0}}}} \right\rangle  ,
\label{average}
\end{aligned}
\end{equation}
where the initial state $|{\psi (0)}\rangle$ is given by Eq.~(\ref{InitialState}). Similarly, we get $\left\langle {{{\hat \sigma }_ + }(0){{\hat c}_{{{cw}}}}(\tau )} \right\rangle $ and $\left\langle {{{\hat \sigma }_ + }(0){{\hat c}_{{{ccw}}}}(\tau )} \right\rangle $. Through straightforward calculations, we derive the following differential equation
%\begin{small}
\begin{equation}
\begin{aligned}
\frac{d}{{d\tau }}\left[ {\begin{array}{*{20}{c}}
  {\left\langle {{{\hat \sigma }_ + }(0){{\hat \sigma }_ - }(\tau )} \right\rangle } \\
  {\left\langle {{{\hat \sigma }_ + }(0){{\hat c}_{{{cw}}}}(\tau )} \right\rangle } \\
  {\left\langle {{{\hat \sigma }_ + }(0){{\hat c}_{{{ccw}}}}(\tau )} \right\rangle }
\end{array}} \right] =  - i{{\mathbf{Q}}_c}\left[ {\begin{array}{*{20}{c}}
  {\left\langle {{{\hat \sigma }_ + }(0){{\hat \sigma }_ - }(\tau )} \right\rangle } \\
  {\left\langle {{{\hat \sigma }_ + }(0){{\hat c}_{{{cw}}}}(\tau )} \right\rangle } \\
  {\left\langle {{{\hat \sigma }_ + }(0){{\hat c}_{{{ccw}}}}(\tau )} \right\rangle }
\end{array}} \right] .
\label{dtau}
\end{aligned}
\end{equation}
%\end{small}
Applying the Laplace transform with the initial conditions $\left\langle {{{\hat \sigma }_ + }(0){{\hat \sigma }_ - }(0)} \right\rangle  = 1$, $\left\langle {{{\hat \sigma }_ + }(0){{\hat c}_{{{ccw}}}}(0)} \right\rangle  = 0$, and $\left\langle {{{\hat \sigma }_ + }(0){{\hat c}_{{{cw}}}}(0)} \right\rangle  = 0$, the spontaneous emission spectrum of the atom can be derived as
\begin{equation}
\begin{aligned}
S(\omega ) = \frac{1}{\pi }\frac{{\Gamma (\omega )}}{{{{[\omega  - {\omega _0} + {\omega _c} - \Delta (\omega )]}^2} + {{(\frac{{\Gamma (\omega )}}{2})}^2}}},
\label{SEspectrum}
\end{aligned}
\end{equation}
where $\Gamma(\omega)=-2 g^{2} \operatorname{Im}[\chi(\omega)]$ is the local coupling strength, while $\Delta(\omega)=g^{2} \operatorname{Re}[\chi(\omega)]$ denotes the photonic Lamb shift, with $\chi(\omega)$ being the response function of the optical cavity
\begin{equation}
\begin{aligned}
\chi (\omega ) = \frac{2}{{\omega  + i\frac{\gamma }{2} + JM}} - \frac{ - 2J + i\gamma {e^{i\phi }}}{{(\omega  + i\frac{\gamma }{2} + JM)({\omega  + i\frac{\gamma }{2}})}},
\label{chiomega}
\end{aligned}
\end{equation}
where $M = {(- J + i\gamma {e^{i\phi }}) }/{(\omega  + i\frac{\gamma }{2})}$.

In the subsequent sections, based on the coupled dynamical equations given by Eq.~(\ref{deltap}), we will discuss the Markovian dynamics for the vacancy-like dressed bound state and Friedrich-Wintgen dressed bound state in Sec.~III. The corresponding non-Markovian form of Eq.~(\ref{deltap}) will be derived in Sec.~IV.

\section{Dressed bound states in Markovian case}

\subsection{Vacancy-Like dressed bound state}

The vacancy-like dressed bound state is a single-photon dressed state having exactly the same energy as the bare atom irrespective of the coupling strength under the rotating wave approximation \cite{Leonforte063601}. The coupled cavity is the simplest model that supports the vacancy-like DBS, where the atom interacts with one of two cavities. Eigenstates with similar properties are frequently mentioned in several fields such as quantum biology \cite{Caruso105106} and dark states in atomic physics \cite{Lambropoulos325}.
%At first glance our model is different from the coupled cavity proposed in Ref.~\cite{Leonforte063601}, but similar vacancy-like dressed bound state can also appear through some transformations.
To find the condition of the vacancy-like DBS in the optical cavity, we express $\hat c_{\mathrm{ccw}}$ and $\hat c_{\mathrm{cw}}$ in terms of the standing-wave modes operators $\hat c_{1}$ and $\hat c_{2}$
\begin{equation}
\begin{aligned}
{{\hat c}_{{{cw}}}} = \frac{1}{{\sqrt 2 }}\left( {{{\hat c}_1} + {{\hat c}_2}} \right),\quad {{\hat c}_{{{ccw}}}} = \frac{1}{{\sqrt 2 }}\left( {{{\hat c}_1} - {{\hat c}_2}} \right).
\label{standingwavemodes}
\end{aligned}
\end{equation}
%Substituting Eq.~(\ref{standingwavemodes}) into Eq.~(\ref{deltap}), we obtain a set of the differential equations for the probability amplitudes
%\begin{equation}
%\begin{aligned}
%\frac{d}{{dt}}\left\langle {{\sigma _ - }} \right\rangle  &=  - i\left[ {({\omega _0} - {\omega _c})\left\langle {{\sigma _ - }} \right\rangle  + \sqrt 2 g\left\langle {{c_1}} \right\rangle } \right],\\
%\frac{d}{{dt}}\left\langle {{c_1}} \right\rangle  &=  - i\left[ {\sqrt 2 g\left\langle {{\sigma _ - }} \right\rangle  + \left( {J - i\frac{{\gamma \left( {1 + {e^{i\phi }}} \right)}}{2}} \right)\left\langle {{c_1}} \right\rangle  + i\frac{\gamma }{2}{e^{i\phi }}\left\langle {{c_2}} \right\rangle } \right],\\
%\frac{d}{{dt}}\left\langle {{c_2}} \right\rangle  &=  - i\left[ { - i\frac{\gamma }{2}{e^{i\phi }}\left\langle {{c_1}} \right\rangle  + \left( { - J - i\frac{{\gamma \left( {1 - {e^{i\phi }}} \right)}}{2}} \right)\left\langle {{c_2}} \right\rangle } \right],
%\label{averageEq}
%\end{aligned}
%\end{equation}

%figure 2
 \begin{figure}[t]
   \centerline{
   \includegraphics[width=8.4cm, height=4cm, clip]{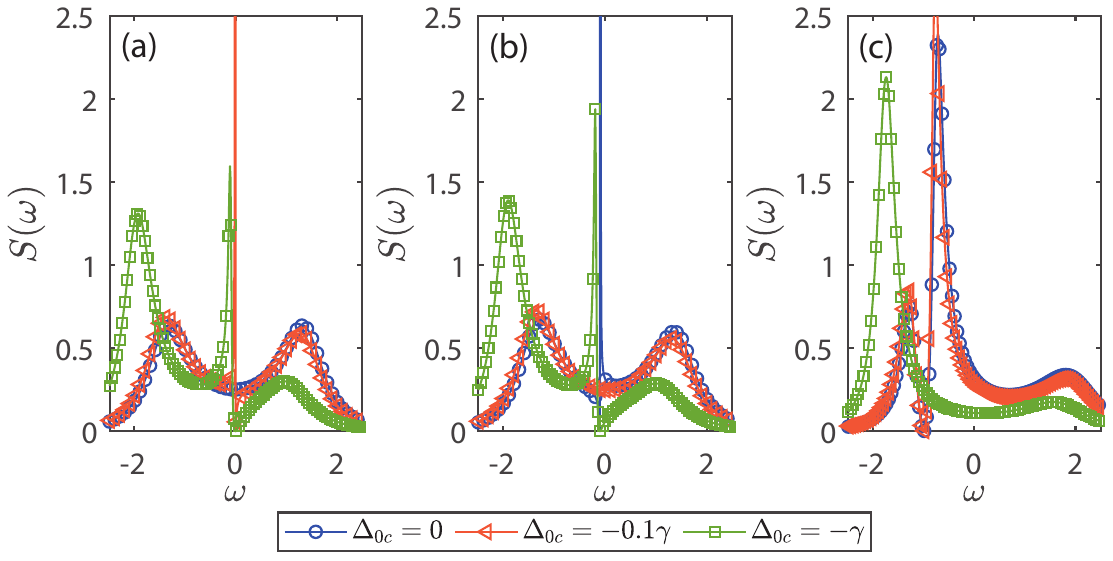}}
   \caption{The spontaneous emission spectrum $S(\omega)$ given by Eq.~(\ref{SEspectrum}) as a function of $\omega$ with the various atom-cavity detuning $\Delta_{0c}=\omega_0-\omega_c$. The parameters chosen are $g=\gamma$, $\phi=2n\pi$, (a) $J=0\gamma$; (b) $J=0.1\gamma$; (c) $J=\gamma$.}\label{fig2}
 \end{figure}
 %figure 3
\begin{figure*}[htbp]
   \centerline{
   \includegraphics[width=16cm, height=8.5cm, clip]{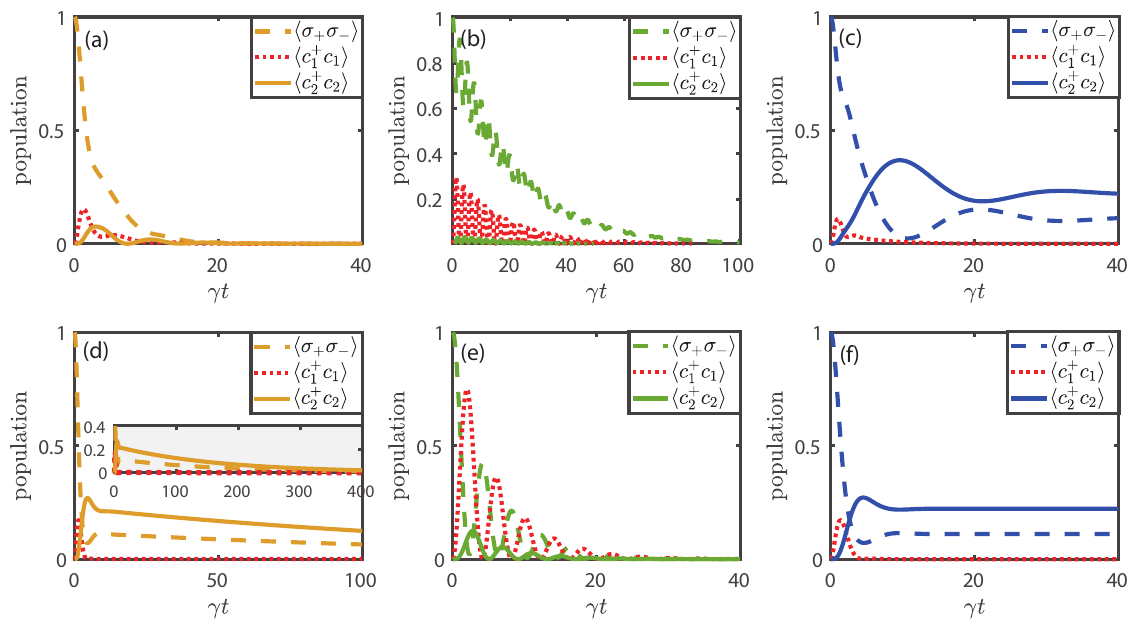}}
   \caption{The evolution of the population in different parameters calculated by Eq.~(\ref{dotABCt}). (a) and (d) only satisfy Eq.~(\ref{MarkVLcondition2}), but not Eq.~(\ref{MarkVLcondition1}). The parameters chosen are $g=0.5\gamma$, $\Delta_{0c}=0$, $\phi=2\pi$, (a) $J=\gamma$, (d) $J=0.1\gamma$. (b) and (e) correspond to meeting Eq.~(\ref{MarkVLcondition1}) instead of Eq.~(\ref{MarkVLcondition2}). The parameters chosen are $g=0.5\gamma$, $\phi=\pi$, (b) $J=-\Delta_{0c}=\gamma$, (e) $J=-\Delta_{0c}=0.1\gamma$. (c) and (f) satisfy the conditions of vacancy-like dressed bound state. The parameters chosen are $g=0.5\gamma$, $\phi=2\pi$, (b) $J=-\Delta_{0c}=\gamma$, (e) $J=-\Delta_{0c}=0.1\gamma$.}\label{fig3}
 \end{figure*}

We obtain $d \vec{\mu} / d t=-i \mathbf{Q}_{s} \vec{\mu}$ with $\vec{\mu}=\left[\left\langle\hat\sigma_{-}\right\rangle,\left\langle\hat c_{1}\right\rangle,\left\langle\hat c_{2}\right\rangle\right]^{T}$, where the matrix $\mathbf{Q}_{s}$ takes the form
\begin{equation}
\begin{aligned}
{{\mathbf{Q}}_s} = \left[ {\begin{array}{*{20}{c}}
  {{\omega _0} - {\omega _c}}&{\sqrt 2 g}&0 \\
  {\sqrt 2 g}&{J - i\frac{{\gamma \left( {1 + {e^{i\phi }}} \right)}}{2}}&{i\frac{\gamma }{2}{e^{i\phi }}} \\
  0&{ - i\frac{\gamma }{2}{e^{i\phi }}}&{ - J - i\frac{{\gamma \left( {1 - {e^{i\phi }}} \right)}}{2}}
\end{array}} \right].
\label{Ms}
\end{aligned}
\end{equation}
In this case, noting that atom is exclusively coupled to $\hat{c}_{1}$ for
\begin{equation}
{\omega _0} - {\omega _c} =  - J \label{MarkVLcondition1}
\end{equation}
and
\begin{equation}
\phi=2 n \pi \,\, (n \in \mathbb{Z}), \label{MarkVLcondition2}
\end{equation}
where the existence of a vacancy-like dressed bound state is guaranteed, whose energy is irrespective of the atom-cavity coupling strength $g$. The vacuum bound state in the rotating frame is $\left| {{\psi _{VL}}} \right\rangle  = {( {{{ - i\gamma }}/{{\sqrt {8{g^2} + {\gamma ^2}} }},0,{{2\sqrt 2 g}}/{{\sqrt {8{g^2} + {\gamma ^2}} }}})^T}$ with eigenvalue $\omega_{VL} = -J$, implying that it is a single-photon dressed state with a node on the atom.

The presence of a vacancy-like DBS can be verified by examining the spectral density $J(\omega)$ of the optical cavity, which is calculated by $\Gamma (\omega ) =  - 2{g^2}\operatorname{Im} [\chi (\omega )] = 2\pi J(\omega )$. Under the initial conditions $\left\langle {{{\hat \sigma }_ + }(0){{\hat \sigma }_ - }(0)} \right\rangle  = 1$, $\left\langle {{{\hat \sigma }_ + }(0){{\hat c}_{{{ccw}}}}(0)} \right\rangle  =  \left\langle {{{\hat \sigma }_ + }(0){{\hat c}_{{{cw}}}}(0)} \right\rangle  = 0$, the spectral density for vacancy-like dressed bound state can be analytically obtained
\begin{equation}
\begin{aligned}
J(\omega ) = \frac{{2{g^2}\gamma {{(\omega  + J)}^2}}}{{\pi \left[ {{{({\omega ^2} - \frac{{{\gamma ^2}}}{4} - {J^2})}^2} + {{\left( {\omega \gamma  - J\gamma } \right)}^2}} \right]}},
\label{Jw}
\end{aligned}
\end{equation}
which shows that the spectral density becomes zero for $\omega=-J=\omega_0-\omega_c$, implying a null electric-field amplitude at the location of the atom. Physically, this signifies that there is no available channel for the atom to decay, consistent with the nature of the vacancy-like DBS.

Figure~\ref{fig2} shows the spontaneous emission spectrum $S(\omega)$ as a function of $\omega$ for various detunings $\Delta_{0c}=\omega_0-\omega_c$, which is solved by Eq.~(\ref{SEspectrum}). In Fig.~\ref{fig2}(a), the spontaneous emission spectrum $S(\omega)$ forms a triplet deviating from the DBS at $\Delta_{0c}=0$ for $J=0$, exhibiting a Fano-type line shape around the cavity resonance. As the atom energy approaches the cavity resonance, the central peak sharpens and rises (see red-triangle line); on resonance ($\omega=-J=\Delta_{0c}=0$) the central peak vanishes (see blue-circle line), implying the formation of a vacancy-like DBS. Moreover, increasing the coupling strength to $J/\gamma=0.1$ and $J/\gamma=1$ has a similar effect in Fig.~\ref{fig2}(a), as shown in Fig.~\ref{fig2}(b) and (c), respectively. As shown by the green-square line in Fig.~\ref{fig2}(c), the central peak of spontaneous emission spectrum at $\omega/\gamma=-J/\gamma=-1$ disappears for $\Delta_{0c}/\gamma=1$.

In Fig.~\ref{fig3}, we plot the population dynamics for different parameters. We take $J/\gamma=1$ in Figs.~\ref{fig3}(a)-(c) and $J/\gamma=0.1$ in Figs.~\ref{fig3}(d)-(f). For $\phi=2\pi$ and $\Delta_{0c}=0$, the collective amplitudes of atom and cavity are damped rapidly as shown in Fig.~\ref{fig3}(a) since $\Delta_{0c} \neq -J$ does not satisfy the condition for vacancy-like dressed bound state, as also reflected in Fig.~\ref{fig3}(d). In addition, adjusting $\phi=\pi$ and $\Delta_{0c}=-J$ yields similar results as shown in Fig.~\ref{fig3}(b) due to $\phi \neq 2n\pi$. Conversely, when we change $\phi=2\pi$ and $\Delta_{0c}=-J$ in Fig.~\ref{fig3}(c), the steady-state population of atom (blue-dashed line) remains non-zero, while the population of $\hat{c}_1$ (red-dotted line) is depleted at the steady state as the eigenstate $\left| {{\psi _{VL}}} \right\rangle$ indicates.

%figure 4
\begin{figure}[t]
   \centerline{
   \includegraphics[width=8.4cm, height=4cm, clip]{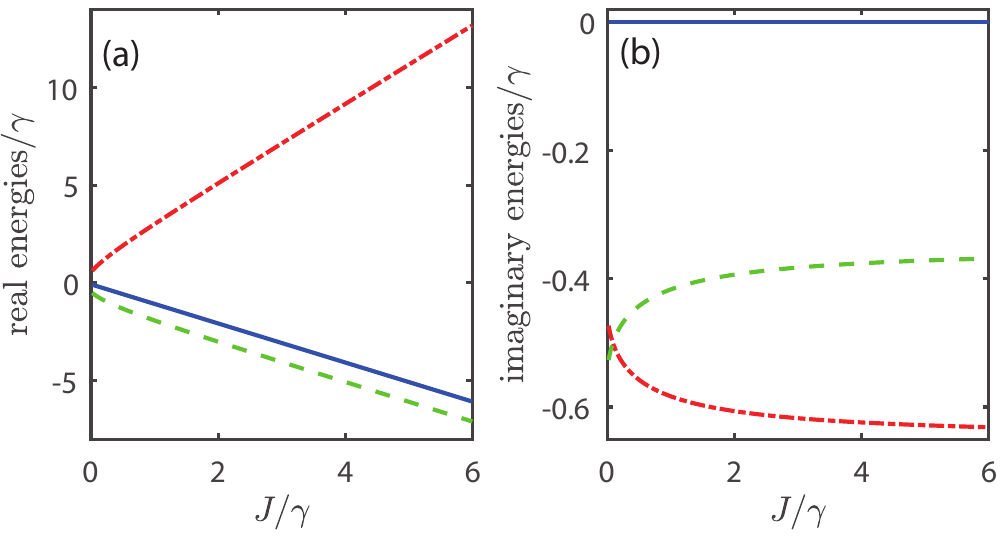}}
   \caption{(a) The real and (b) imaginary parts of eigenenergies corresponding to $\mathbf{Q}_{c}$ shown in Eq.~(\ref{Mc}) versus $J/\gamma$ for the optical cavity with the Friedrich-Wintgen dressed bound state under the condition~(\ref{gfw}). Different curves represent different eigenenergies. The parameters chosen are $\phi=0.1\pi$ and $\Delta_{0c}=0$.}\label{fig4}
 \end{figure}

\subsection{Friedrich-Wintgen dressed bound state}

In addition to the vacancy-like DBS, the optical cavity also supports another type of DBS, which operates via a mechanism akin to the Friedrich-Wintgen bound states in the continuum (BICs) \cite{Marinica183902,Cotrufo799,Hsu16048,Hu1353,Roccati565,Liu7405}. When two resonances reach a degeneracy point by tuning a continuous parameter, interference can induce an avoided level crossing of the frequencies, potentially forming a BIC with a vanishing resonance width at a specific parameter value.

Similar to BICs, our aim is to derive the condition for the existence of the Friedrich-Wintgen DBS in the optical cavity. The matrix $\mathbf{Q}_{c}$ in Eq.~(\ref{Mc}) can be written as follows
\begin{equation}
\begin{aligned}
\mathbf{Q}_{c}=H_{+}-i H_{-},
\label{recastMc}
\end{aligned}
\end{equation}
with
\begin{equation}
\begin{aligned}
H_{-}  = {D^\dag }D,
\label{DgaggerD}
\end{aligned}
\end{equation}
where $H_{+}$ is the Hermitian part giving rise to real energy for the DBS
\begin{equation}
\begin{aligned}
H_{+} = \left( {\begin{array}{*{20}{c}}
  {{\omega _0} - {\omega _c}}&g&g \\
  g&0&{J - i\frac{\gamma }{2}{e^{i\phi }}} \\
  g&{J + i\frac{\gamma }{2}{e^{ - i\phi }}}&0
\end{array}} \right),
\label{HB}
\end{aligned}
\end{equation}
and the dissipative operator governs the imaginary part of the eigenenergies
\begin{equation}
\begin{aligned}
H_{-}  = \left( {\begin{array}{*{20}{c}}
  0&0&0 \\
  0&{\frac{\gamma }{2}}&{\frac{\gamma }{2}{e^{i\phi }}} \\
  0&{\frac{\gamma }{2}{e^{ - i\phi }}}&{\frac{\gamma }{2}}
\end{array}} \right).
\label{gamma}
\end{aligned}
\end{equation}
To be specific, we can calculate the coupling matrix $D = \left( {0,\sqrt {\gamma /2} {e^{ - i\phi }},\sqrt {\gamma /2} } \right)$. A zero eigenvalue in the coupling matrix $D$ implies the existence of a null vector $\left| {{\psi _0}} \right\rangle$ that satisfies
\begin{equation}
\begin{aligned}
D\left|\psi_{0}\right\rangle=0.
\label{anullvector}
\end{aligned}
\end{equation}
%figure 5
\begin{figure}[t]
   \centerline{
   \includegraphics[width=5cm, height=4.5cm, clip]{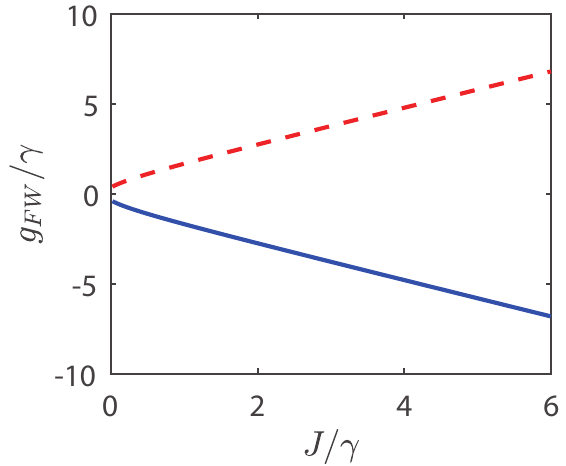}}
   \caption{Condition of the Friedrich-Wintgen dressed bound state versus $J/\gamma$ calculated by Eq.~(\ref{gfw}). The parameters chosen are the same as Fig.~{\ref{fig4}}.}\label{fig5}
 \end{figure}

%figure 6
 \begin{figure}[b]
   \centerline{
   \includegraphics[width=8.4cm, height=6.4cm, clip]{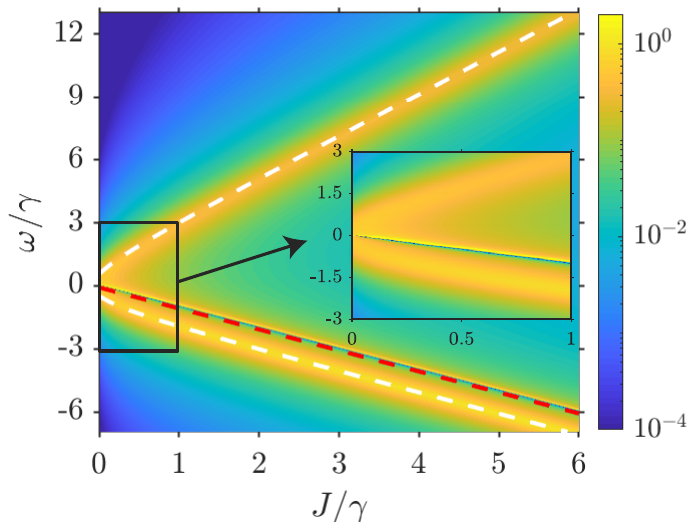}}
   \caption{The spontaneous emission spectrum $S(\omega)$ for the optical cavity with the Friedrich-Wintgen dressed bound state obtained by Eq.~(\ref{SEspectrum}) as a function of $\omega$ and $J/\gamma$. The parameters chosen are the same as in Fig.~{\ref{fig4}}. The white dashed lines track the real eigenenergies, while the red dashed line corresponds to the condition of the Friedrich-Wintgen dressed bound state, which is consistent with Fig.~{\ref{fig4}}. The inset shows the drawing of partial enlargement.}\label{fig4inset}
 \end{figure}

Through straightforward calculations, we obtain $|\psi_{0}\rangle=(\alpha,-e^{-i \phi}, 1)^{T}$, where $\alpha$ is an undetermined coefficient. The Friedrich-Wintgen DBS manifests  when $|\psi_{0}\rangle$ fulfills $H_{+}|\psi_{0}\rangle=\omega_{\mathrm{FW}}|\psi_{0}\rangle$. The solutions yield the energy and condition of the Friedrich-Wintgen DBS
\begin{equation}
\begin{aligned}
{\omega _{{{FW}}}} =   - J - \frac{1}{2}\gamma \tan (\phi /2),
\label{omegafw}
\end{aligned}
\end{equation}
\begin{equation}
\begin{aligned}
{g_{{{FW}}}} =   \pm \frac{{\sqrt {\sin (\frac{\phi }{2})[\vartheta \cos (\frac{\phi }{2}) + \gamma \sin (\frac{\phi }{2})][\gamma  + 2J\sin (\phi )]} }}{{\sqrt 2 \sin (\phi )}},
\label{gfw}
\end{aligned}
\end{equation}
where $\vartheta  = 2(J + {\omega _0} - {\omega _c})$.

%figure 7
\begin{figure}[thpb]
   \centerline{
   \includegraphics[width=8.4cm, height=6.2cm, clip]{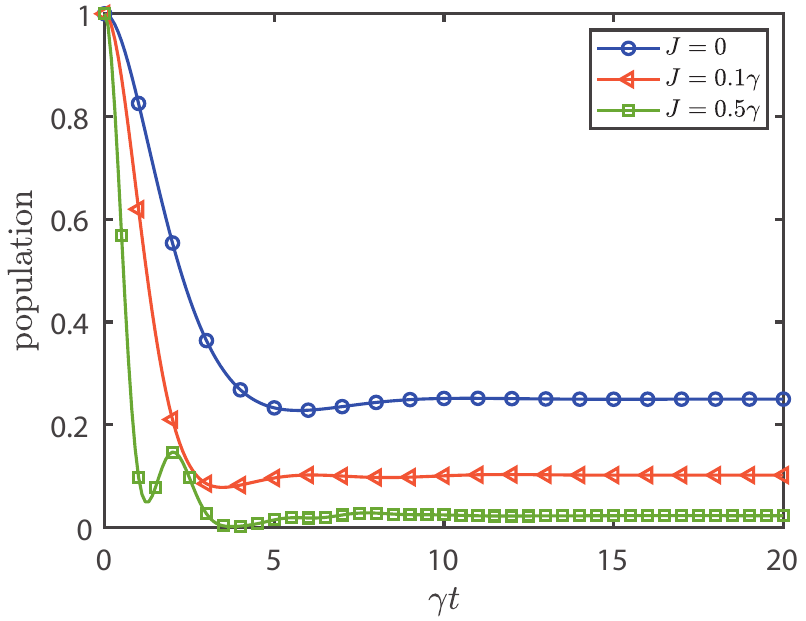}}
   \caption{Evolution of the excited-state population in different $J$ obtained by numerically solving Eq.~(\ref{dotABCt}) for the condition of Friedrich-Wintgen dressed bound state. The parameters chosen are the same as Fig.~{\ref{fig4}}.}\label{fig6}
\end{figure}

Figure~\ref{fig4} respectively plots (a) the real and (b) imaginary parts of eigenenergies versus $J$ for the optical cavity featuring the Friedrich-Wintgen dressed bound state, where the different curves represent different eigenenergies. The imaginary part of the eigenvalue corresponding to the blue line in Fig.~\ref{fig4}(b) remains zero as $J$ varies, indicating the eigenenergy of the Friedrich-Wintgen DBS given by Eq.~(\ref{omegafw}). Moreover, Fig.~\ref{fig5} plots condition $g_{FW}$ of the Friedrich-Wintgen dressed bound state versus $J$ solved by Eq.~(\ref{gfw}). It is evident that $g_{FW+}$ ($g_{FW-}$) increases (decreases) linearly with the increase of $J$.

Figure~\ref{fig4inset} illustrates the spontaneous emission spectrum $S(\omega)$ for the optical cavity with the Friedrich-Wintgen dressed bound state as a function of $\omega$ and $J/\gamma$, which is calculated by Eq.~(\ref{SEspectrum}). The white-dashed lines denote the real parts of the eigenenergies of non-Friedrich-Wintgen BDS, while the red-dashed line corresponds to the real part of the eigenenergy of the Friedrich-Wintgen dressed bound state, which is consistent with Fig.~\ref{fig4}. With $g_{FW}$, only two peaks are observed in the spontaneous emission spectrum. The Rabi peak associated with the Friedrich-Wintgen DBS is invisible due to the vanishing linewidth, as shown in Fig.~\ref{fig4inset}.

Furthermore, we plot time evolution of the excited-state population with various $J$ in Fig.~\ref{fig6}. The different lines correspond to different values $J = 0$ (blue-circle line), $J=0.1\gamma$ (red-triangle line), and $J=0.5\gamma$ (green-square line). With all other parameters fixed, we observe that as time progresses, the population tends to stabilize, implying the formation of the Friedrich-Wintgen dressed bound state. Moverover, we find the steady-state population of Friedrich-Wintgen DBS decreases with the further increase of $J$, which is comprehensible. Since $g_{FW+}$ is proportional to $J$, an increase in $J$ leads to stronger atom-cavity coupling, resulting in a decreased steady-state population of the Friedrich-Wintgen DBS due to more photons being coupled into the optical cavity.

\section{Dressed bound states in non-Markovian systems}

\subsection{Non-Markovian model and dynamics}

In this section, we examine the scenario where dissipation is non-Markovian, characterized by an environment consisting of a continuum of bosonic modes. The cavity interacts with the $k$th mode (eigenfrequency $\omega_k$) of the non-Markovian environment via annihilation and creation operators of this system \cite{Liu931,Groblacher7606,Xiong032101,Cialdi052104,Khurana022107,Madsen233601,Guo230401,Li140501,Hoeppe043603,Xu042328,Tang10002,Uriri052107,Anderson3202,Liu062208}. By transitioning to a rotating frame defined by the unitary transformation $U=\exp [i(\omega_c \sigma_+ \sigma_- + {\omega _c}\hat{c}_{cw}^\dag {\hat{c}_{cw}} + {\omega _c}\hat{c}_{ccw}^\dag {\hat{c}_{ccw}} + \int {d\omega \omega b_k^{\dag} b_k})t]$, the total Hamiltonian~(\ref{HamiltonianTotal}) is changed to
\begin{align}
   \hat{H}_{NM} =& {\hat{H}_S}+ {\hat{H}_{SB}},\nonumber\\
  {\hat{H}_S} =& {\Delta _{0c}}{\hat{\sigma} _ + }{\hat{\sigma} _ - } + J(\hat{c}_{cw}^\dag {\hat{c}_{ccw}} + \hat{c}_{ccw}^\dag {\hat{c}_{cw}})\nonumber\\
               &+ \sum\limits_{j = cw,ccw} {g(\hat{c}_j^\dag {\hat{\sigma} _ - } + {\hat{\sigma} _ + }{\hat{c}_j})},\label{HamiltonianNonMarkovian}\\
  {\hat{H}_{SB}} =& i\sum\limits_{j = cw,ccw} {\int {d\omega \gamma_k\hat{b}_k^\dag {e^{ - ik{x_j}}}{\hat{c}_j}{e^{ - i(\omega-\omega_c)t}}}  + \rm{H.c.}},\nonumber
\end{align}
where $\gamma_k$ denotes the coupling strength between the CCW (CW) mode and the environment with frequency $\omega $.
Hamiltonian~(\ref{HamiltonianNonMarkovian}) is analytically solvable due to the conservation of the total excitation number $\hat{N}={\hat{\sigma}_+}{\hat{\sigma}_-} + \hat{c}_{\mathrm{cw}}^\dag {\hat{c}_{\mathrm{cw}}} + \hat{c}_{\mathrm{ccw}}^\dag {\hat{c}_{\mathrm{ccw}}} + \int {d\omega \hat{b}_k^\dag {\hat{b}_k} }$. We assume that the initial state is prepared in $  | {\psi (0)} \rangle  = | {e,0,0,{0}} \rangle$, where $| {e,0,0,{0}} \rangle = | e \rangle \otimes | 0 \rangle_{\mathrm{cw}} \otimes | 0 \rangle_{\mathrm{ccw}} \otimes | 0 \rangle $. $| 0 \rangle$ denotes that all the modes of the environment are in the vacuum state represented by $| 0 \rangle = |0_1\rangle \otimes |0_2\rangle \ldots \otimes |0_\infty \rangle$. The state of the total system can be expressed as
\begin{equation}
\begin{aligned}
\left| {\psi (t)} \right\rangle  = &{C_0}\left| {{{g,0,0,}}{{{0}}}} \right\rangle  + A(t)\left| {{{e,0,0,}}{{{0}}}} \right\rangle  + B(t)\left| {{{g,1,0,}}{{{0}}}} \right\rangle  \\
&+ C(t)\left| {{{g,0,1,}}{{{0}}}} \right\rangle  + \int {d\omega D_k(t)\hat{b}_k^\dag  \left| {{{g,0,0,}}{{{0}}}} \right\rangle },
\label{expressstateNonMarkovian}
\end{aligned}
\end{equation}
where $\hat{b}_k^\dag  \left| {{{g,0,0,}}{{{0}}}} \right\rangle$ denotes the waveguide with only a single excitation in the $k$th field mode with the probability $|D_k(t)|^2$. Substituting Eq.~(\ref{expressstateNonMarkovian}) into the Schr\"{o}dinger equation $i|\dot{\psi}(t)\rangle=\hat{H}_{NM}|\psi(t)\rangle$, we derive a set of differential equations for the probability amplitudes
\begin{subequations}
\begin{align}
i\dot A(t) = &({\omega _0} - {\omega _c})A(t) + gB(t) + gC(t),\label{dotdtNonMarkovian1}\\
i\dot B(t) = &gA(t) + JC(t) - i\int {d\omega {\gamma_k^*}} {e^{ - i(\omega  - {\omega _c})t}}{e^{ik{x_{c\omega }}}}{D_k}(t),\label{dotdtNonMarkovian2}\\
i\dot C(t) = &gA(t) + JB(t) - i\int {d\omega {\gamma_k^*}} {e^{ - i(\omega  - {\omega _c})t}}{e^{ik{x_{cc\omega }}}}{D_k}(t),\label{dotdtNonMarkovian3}\\
i{{\dot D}_k}(t) = &i\gamma_k{e^{i(\omega  - {\omega _c})t}}[{e^{ - ik{x_{c\omega }}}}B(t) + {e^{ - ik{x_{cc\omega }}}}C(t)] .\label{dotdtNonMarkovian4}
\end{align}
\end{subequations}
Solving Eq.~(\ref{dotdtNonMarkovian4}), we obtain the solution of the environment probability amplitude
\begin{equation}
\begin{aligned}
{D_k}(t) = &\int_0^t {\gamma_k} {e^{i(\omega  - {\omega _c})t}}{e^{ - ik{x_{c\omega }}}}B(\tau )d\tau  \\
&+ \int_0^t {\gamma_k{e^{i(\omega  - {\omega _c})t}}{e^{ - ik{x_{cc\omega }}}}C(\tau )} d\tau .
\label{dotdtjfNonMarkovian}
\end{aligned}
\end{equation}
The first and the second terms on the right-hand side of Eq.~(\ref{dotdtjfNonMarkovian}) describes the influence of the CW and CCW modes on ${D_k}(t)$, respectively.
Eqs.~(\ref{dotdtNonMarkovian2})-~(\ref{dotdtNonMarkovian3}) can be rewritten as a set of integro-differential equations for the cavity probability amplitude by substituting Eq.~(\ref{dotdtjfNonMarkovian}) into Eqs.~(\ref{dotdtNonMarkovian2}) and~(\ref{dotdtNonMarkovian3})
\begin{equation}
\begin{aligned}
i\dot A(t) = &{\Delta _{0c}}A(t) + gB(t) + gC(t),\\
i\dot B(t) = &gA(t) + JC(t) - i\int_0^t {d\tau } {f_1}(t - \tau )B(\tau )\\
& - i\int_0^t {d\tau {f_2}(t - \tau )C(\tau )},\\
i\dot C(t) = &gA(t) + JB(t) - i\int_0^t {d\tau } {f_3}(t - \tau )B(\tau )\\
&- i\int_0^t {d\tau {f_1}(t - \tau )C(\tau )}   ,
\label{dotABCtjfNonMarkovian}
\end{aligned}
\end{equation}
where ${f_1}(t - \tau ) = \int {d\omega } {\left| {\gamma_k} \right|^2}{e^{ - i(\omega  - {\omega _c})(t - \tau )}}$, ${f_2}(t - \tau ) = \int {d\omega } {\left| {\gamma_k} \right|^2}{e^{ - i(\omega  - {\omega _c})(t - \tau )}}{e^{ik({x_{c\omega }} - {x_{cc\omega }})}}$, and ${f_3}(t - \tau ) = \int {d\omega } {\left| {\gamma_k} \right|^2}{e^{ - i(\omega  - {\omega _c})(t - \tau )}}{e^{ - ik({x_{c\omega }} - {x_{cc\omega }})}}$.
The correlation function is given by ${f_2}(t - \tau ) = {e^{i{\omega _c}{x_1}/\upsilon }}\int {d\omega } {\left| {\gamma_k} \right|^2}{e^{ - i(\omega  - {\omega _c})(t - \tau  - {x_1}/\upsilon )}}$ and ${f_3}(t - \tau ) = {e^{ - i{\omega _c}{x_1}/\upsilon }}\int {d\omega } {\left| {\gamma_k} \right|^2}{e^{ - i(\omega  - {\omega _c})(t - \tau  + {x_1}/\upsilon )}}$, which describe the non-Markovian fluctuation-dissipation relationship of environment. ${x_1} = {x_{c\omega }} - {x_{cc\omega }}$ represents the relative locations of the CCW mode and the mirrored CW mode. We define the Lorentzian coupling strength as \cite{Xiong032107,Shen012156,Uhlenbeck823,Gillespie2084,Jing240403}
\begin{equation}
\begin{aligned}
\gamma_k = \sqrt {\frac{{{\gamma}}}{{2\pi }}} \frac{\lambda }{{\lambda  - i(\omega  - {\omega _c})}},
\label{Lorentzianspectraldensity}
\end{aligned}
\end{equation}
where $\lambda$ is the non-Markovian environmental spectrum width, while ${\gamma}$ denotes the dissipation strength between WGM and waveguide. With Eq.~(\ref{Lorentzianspectraldensity}), we get
\begin{equation}
\begin{aligned}
{f_1}(t - \tau ) &= \frac{1}{2}{e^{ - \lambda \left| {t - \tau } \right|}}{\gamma}\lambda ,\\
{f_2}(t - \tau ) &= \frac{1}{2}{e^{i\phi }}{e^{ - \lambda \left| {t - \tau  - {x_1}/\upsilon } \right|}}{\gamma}\lambda ,\\
{f_3}(t - \tau ) &= \frac{1}{2}{e^{ - i\phi }}{e^{ - \lambda \left| {t - \tau  + {x_1}/\upsilon } \right|}}{\gamma}\lambda ,
\label{f123}
\end{aligned}
\end{equation}
where $\phi  = {\omega _c}{x_1}/\upsilon $.
For convenience, we define $X = \int_0^t {d\tau {f_1}(t - \tau )B(\tau )}$, $Y = \int_0^t {d\tau {f_2}(t - \tau )C(\tau )}$ and $Z = \int_0^t {d\tau {f_1}(t - \tau )C(\tau )}  + \int_0^t {d\tau {f_3}(t - \tau )B(\tau )}$. Through simple calculation, we obtain $\vec \varepsilon = {\left[ {\left\langle {{{\hat \sigma }_ - }} \right\rangle ,\left\langle {{{\hat c}_{{{ccw}}}}} \right\rangle ,\left\langle {{{\hat c}_{{{cw}}}}} \right\rangle ,{C_0}^*X(t),{C_0}^*Y(t),{C_0}^*Z(t)} \right]^T}$, and the matrix $\mathbf{Q}_{c}$ being
%figure 8
\begin{figure}[b]
   \centerline{
   \includegraphics[width=8.4cm, height=6.2cm, clip]{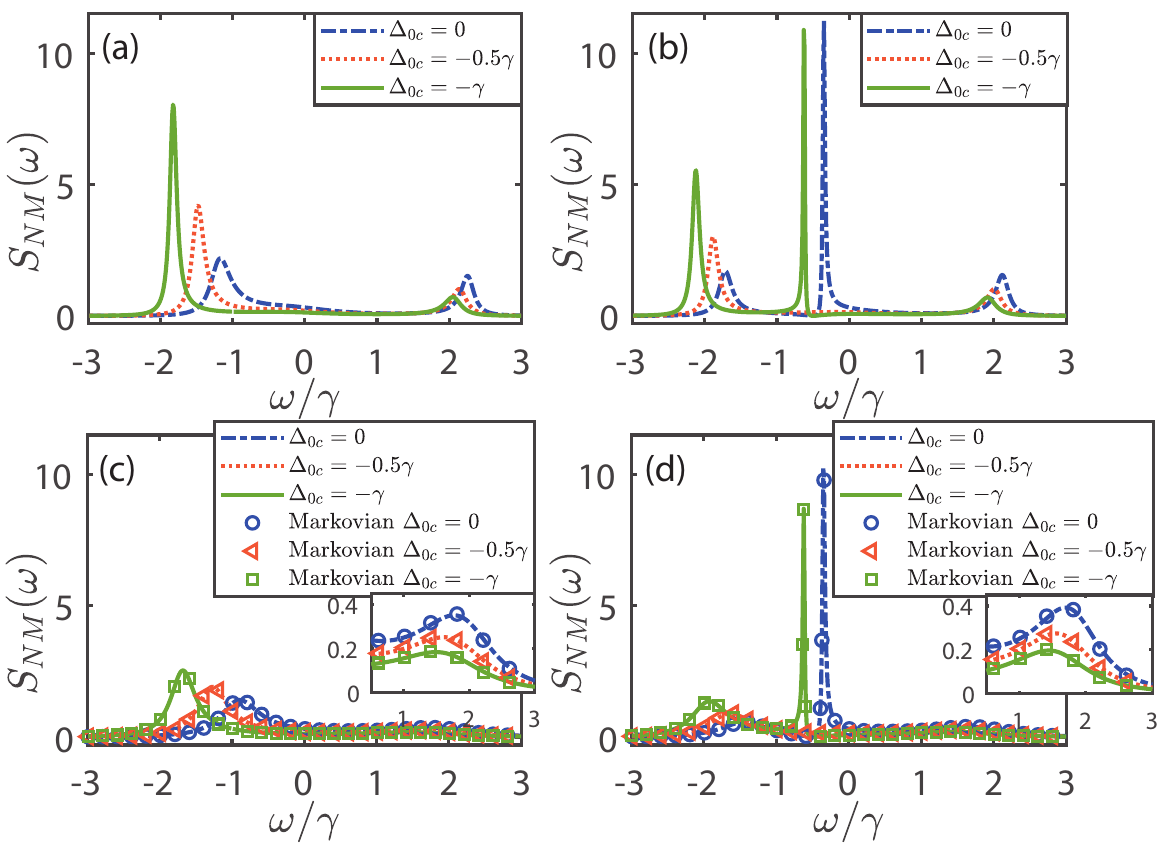}}
   \caption{(a)-(b) Non-Markovian spontaneous emission spectrum $S_{NM}(\omega)$ obtained by Eq.~(\ref{SEspectrumNonMarkovian}) as a function of $\omega$ with various the atom-cavity detuning $\Delta_{0c}=\omega_0-\omega_c$. The parameters chosen are $g=\gamma$, $J=\gamma$, $\lambda=\gamma$, (a) $\theta=0$; (b) $\theta=\pi/3$. (c)-(d) Non-Markovian spontaneous emission spectrum $S_{NM}(\omega)$ for the case of the Markovian limit (different styles of curves) calculated by Eq.~(\ref{SEspectrumNonMarkovian}) and Markovian spontaneous emission spectrum (different styles of data point symbols) given by Eq.~(\ref{SEspectrumMarkovianspecial}) as a function of $\omega$ with various the atom-cavity detuning $\Delta_{0c}$. The parameters chosen are $\lambda=50\gamma$, (c) $\theta=0$; (d) $\theta=\pi/3$.  The other parameters are the same as in (a).}\label{nMfig3}
 \end{figure}

\begin{widetext}
\begin{equation*}
\begin{aligned}
{{\mathbf{Q}}_{NM}} = \left[ {\begin{array}{*{20}{c}}
  { - i({\omega _0} - {\omega _c})}&{ - ig}&{ - ig}&0&0&0 \\
  { - ig}&0&{ - iJ}&0&0&{ - 1} \\
  { - ig}&{ - iJ}&0&{ - 1}&{ - 1}&0 \\
  0&0&{\frac{1}{2}{\gamma}\lambda }&{ - \lambda }&0&0 \\
  0&{\frac{1}{2}{e^{i\phi }}{e^{ - \lambda {x_1}/\upsilon }}}&0&0&{\lambda \Theta ({x_1}/\upsilon  - t) - \lambda \operatorname{sgn} (t - \tau  - {x_1}/\upsilon )\Theta (t - {x_1}/\upsilon )}&0 \\
  0&{\frac{1}{2}{\gamma}\lambda }&{\frac{1}{2}{e^{ - i\phi }}{e^{ - \lambda {x_1}/\upsilon }}{\gamma}\lambda }&0&0&{ - \lambda }
\end{array}} \right]   ,
\label{McNonMarkovianordinary}
\end{aligned}
\end{equation*}
\end{widetext}
where $\Theta (x) $ is the unit step function, while $\operatorname{sgn} (x)$ represents the sign function.

\subsection{The case of zero accumulated phase factor}

For the convenience of discussion, we first consider the simplest case ${x_{cw}} - {x_{ccw}} = 0$, which implies that the correlation functions are identical, i.e.  ${f_1}(t - \tau ) = {f_2}(t - \tau ) = {f_3}(t - \tau ) = \frac{1}{2}{e^{ - \lambda \left| {t - \tau } \right|}}{\gamma}\lambda$. The set of the differential equations for the probability amplitudes in Eq.~(\ref{dotABCtjfNonMarkovian}) can be reduced to
\begin{equation}
\begin{aligned}
i\dot A(t) &= ({\omega _0} - {\omega _c})A(t) + gB(t) + gC(t) , \\
i\dot B(t) &= gA(t) + JC(t) - iW(t)  ,\\
i\dot C(t) &= gA(t) + JB(t) - iW(t) ,\\
\dot W(t) &=  - \lambda W(t) + \frac{1}{2}{\gamma}\lambda B(t) + \frac{1}{2}{\gamma}\lambda C(t)  ,
\label{dotABCtNonMarkovian}
\end{aligned}
\end{equation}
where we have defined $ W(t) = \int_0^t {d\tau {f_1}(t - \tau )} B(\tau ) + \int_0^t {d\tau {f_1}(t - \tau )} C(\tau )$. Details of the derivation for Eq.~(\ref{dotABCtNonMarkovian}) can be found in Appendix \ref{APPB}. To incorporate the effects of non-Hermitian interaction, we make the replacement $J \to J{e^{i\theta }}$ for Eq.~(\ref{HamiltonianNonMarkovian}) as $J{e^{i\theta }}\hat c_{{\rm{cw}}}^\dag {\hat c_{{\rm{ccw}}}} + J{e^{ - i\theta }}\hat c_{{\rm{ccw}}}^\dag {\hat c_{{\rm{cw}}}}$ and takes $\vec \varepsilon = {\left[ {\left\langle {{{\hat \sigma }_ - }} \right\rangle ,\left\langle {{{\hat c}_{{{ccw}}}}} \right\rangle ,\left\langle {{{\hat c}_{{{cw}}}}} \right\rangle ,W(t)} \right]^T}$. The matrix $\mathbf{Q}_{c}$ is
%\begin{equation}
%\begin{aligned}
%{{\mathbf{Q}}_c} = \left[ {\begin{array}{*{20}{c}}
%  {{\omega _0} - {\omega _c}}&g&g&0 \\
%  g&0&{J}&{ - i} \\
%  g&{J}&0&{ - i} \\
%  0&{\frac{i}{2}{\gamma}\lambda }&{\frac{i}{2}{\gamma}\lambda }&{ - i\lambda }
%\end{array}} \right],
%\label{McNonMarkovian}
%\end{aligned}
%\end{equation}
\begin{equation}
\begin{aligned}
{{\mathbf{Q}}_{NMc}} = \left[ {\begin{array}{*{20}{c}}
  {{\omega _0} - {\omega _c}}&g&g&0 \\
  g&0&{J{e^{i\theta }}}&{ - i} \\
  g&{J{e^{ - i\theta }}}&0&{ - i} \\
  0&{\frac{i}{2}{\gamma}\lambda }&{\frac{i}{2}{\gamma}\lambda }&{ - i\lambda }
\end{array}} \right],
\label{McNonMarkovian}
\end{aligned}
\end{equation}
After straightforward calculations, we obtain the differential equation of the non-Markovian system as follows:
\begin{equation}
\begin{aligned}
\frac{d}{{d\tau }}\left[ {\begin{array}{*{20}{c}}
  {\left\langle {{{\hat \sigma }_ + }(0){{\hat \sigma }_ - }(\tau )} \right\rangle } \\
  {\left\langle {{{\hat \sigma }_ + }(0){{\hat c}_{{{cw}}}}(\tau )} \right\rangle } \\
  {\left\langle {{{\hat \sigma }_ + }(0){{\hat c}_{{{ccw}}}}(\tau )} \right\rangle } \\
  {W(\tau )}
\end{array}} \right] =  - i{{\mathbf{Q}}_c}\left[ {\begin{array}{*{20}{c}}
  {\left\langle {{{\hat \sigma }_ + }(0){{\hat \sigma }_ - }(\tau )} \right\rangle } \\
  {\left\langle {{{\hat \sigma }_ + }(0){{\hat c}_{{{cw}}}}(\tau )} \right\rangle } \\
  {\left\langle {{{\hat \sigma }_ + }(0){{\hat c}_{{{ccw}}}}(\tau )} \right\rangle } \\
  {W(\tau )}
\end{array}} \right].
\label{dtauNonMarkovian}
\end{aligned}
\end{equation}
Applying the Laplace transform with the initial conditions $\left\langle {{{\hat \sigma }_ + }(0){{\hat \sigma }_ - }(0)} \right\rangle  = 1$, $\left\langle {{{\hat \sigma }_ + }(0){{\hat c}_{{{ccw}}}}(0)} \right\rangle  = 0$, and $\left\langle {{{\hat \sigma }_ + }(0){{\hat c}_{{{cw}}}}(0)} \right\rangle  = 0$ in the non-Markovian case, the spontaneous emission spectrum of the atom can be derived as
\begin{equation}
\begin{aligned}
S_{NM}(\omega ) = \frac{1}{\pi }\frac{{\Gamma_{NM} (\omega )}}{{{{\left[ {\omega  - {\omega _0} + {\omega _c} - \Delta_{NM} (\omega )} \right]}^2} + {{\left[ {\frac{{\Gamma_{NM} (\omega )}}{2}} \right]}^2}}},
\label{SEspectrumNonMarkovian}
\end{aligned}
\end{equation}
where $\Gamma_{NM}(\omega)=-2 g^{2} \operatorname{Im}[\chi_{NM}(\omega)]$ is the local coupling strength while $\Delta_{NM}(\omega)=g^{2} \operatorname{Re}[\chi_{NM}(\omega)]$ denotes the photonic Lamb shift with $\chi_{NM}(\omega)$ being the response function of the optical cavity
\begin{small}
\begin{equation}
\begin{aligned}
\chi_{NM} (\omega ) =  \frac{{ - 2( - i\omega  + \lambda )(J + J{e^{2i\theta }} + 2\omega {e^{i\theta }})}}{{2{e^{i\theta }}\left[ {( - i\omega  + \lambda )({J^2} - {\omega ^2}) - i\omega \gamma \lambda } \right] - iJ\gamma \lambda (1 + {e^{2i\theta }})}}.
\label{chiomegaNonMarkovian}
\end{aligned}
\end{equation}
\end{small}

In the Markovian case, we obtain
\begin{equation}
\begin{aligned}
{{\mathbf{Q}}_{M}} = \left[ {\begin{array}{*{20}{c}}
  {{\omega _0} - {\omega _c}}&g&g \\
  g&{ - i\frac{\gamma }{2}}&{J{e^{i\theta}} - i\frac{\gamma }{2}} \\
  g&{J{e^{-i\theta }} - i\frac{\gamma }{2}}&{ - i\frac{\gamma }{2}}
\end{array}} \right],
\label{McMarkovianspecial}
\end{aligned}
\end{equation}
for the zero accumulated phase factor ${x_{cw}} - {x_{ccw}} = 0$, which is defferent from Eq.~({\ref{Mc}}) for the non-zero accumulated phase factor. The non-Markovian matrix~({\ref{McNonMarkovian}}) returns back to the system described by Eq.~(\ref{McMarkovianspecial}) in the Markovian approximation ($\lambda \to \infty$). The spontaneous emission spectrum of the atom can be derived as
\begin{equation}
\begin{aligned}
S_{M}(\omega ) = \frac{1}{\pi }\frac{{\Gamma_{M} (\omega )}}{{{{\left[ {\omega  - {\omega _0} + {\omega _c} - \Delta_{M} (\omega )} \right]}^2} + {{\left[ {\frac{{\Gamma_{M} (\omega )}}{2}} \right]}^2}}},
\label{SEspectrumMarkovianspecial}
\end{aligned}
\end{equation}
where $\Gamma_{M}(\omega)=-2 g^{2} \operatorname{Im}[\chi_{M}(\omega)]$ is the local coupling strength, while $\Delta_{M}(\omega)=g^{2} \operatorname{Re}[\chi_{M}(\omega)]$ denotes the photonic Lamb shift  with $\chi_{M}(\omega)$ being the response function of the optical cavity
\begin{equation}
\begin{aligned}
\chi_{M} (\omega ) = \frac{{2(J + J{e^{2i\theta }} + 2\omega {e^{i\theta }})}}{{iJ\gamma  + iJ\gamma {e^{2i\theta }} - 2{e^{i\theta }}({J^2} - {\omega ^2} - i\omega \gamma )}}.
\label{chiomegaMarkovianspecial}
\end{aligned}
\end{equation}

In the non-Markovian case, we rewrite $\hat c_{\mathrm{ccw}}$ and $\hat c_{\mathrm{cw}}$ in terms of the operators that represent the standing-wave modes $\hat c_{1}$ and $\hat c_{2}$
\begin{equation}
\begin{aligned}
{{\hat c}_{{{cw}}}} = \frac{1}{{\sqrt 2 }}\left( {{{\hat c}_1} + {{\hat c}_2}} \right),\quad {{\hat c}_{{{ccw}}}} = \frac{1}{{\sqrt 2 }}\left( {{{\hat c}_1} - {{\hat c}_2}} \right).
\label{standingwavemodesnonMarkovian}
\end{aligned}
\end{equation}
We obtain $d \vec{\mu} / d t=-i \mathbf{Q}_{NMs} \vec{\mu}$, with $\vec \mu = {\left[ {\left\langle {{{\hat \sigma }_ - }} \right\rangle ,\left\langle {{{\hat c}_1}} \right\rangle ,\left\langle {{{\hat c}_2}} \right\rangle ,W(t)} \right]^T}$. The matrix $\mathbf{Q}_{NMs}$ takes the form
\begin{equation}
\begin{aligned}
{{\mathbf{Q}}_{NMs}} = \left[ {\begin{array}{*{20}{c}}
  {{\omega _0} - {\omega _c}}&{\sqrt 2 g}&0&0 \\
  {\sqrt 2 g}&{J\cos \theta }&{ - iJ\sin \theta }&{ - i\sqrt 2 } \\
  0&{iJ\sin \theta }&{ - J\cos \theta }&0 \\
  0&{\frac{i}{{\sqrt 2 }}{\gamma}\lambda }&0&{ - i\lambda }
\end{array}} \right],
\label{MsvdbsnonMarkovian}
\end{aligned}
\end{equation}
when
\begin{equation}
\begin{aligned}
{\omega _0} - {\omega _c} =  - J\cos \theta,
\label{nonMarkovianVLcondition}
\end{aligned}
\end{equation}
implying exists eigenvalue $ - J\cos \theta $ and eigenvector ${\left[ {{{iJ\sin \theta } \mathord{\left/
 {\vphantom {{iJ\sin \theta } {\sqrt 2 g}}} \right.
 \kern-\nulldelimiterspace} {\sqrt 2 g}},0,1,0} \right]^T}$.

%figure 9
  \begin{figure}[t]
   \centerline{
   \includegraphics[width=8.6cm, height=6.6cm, clip]{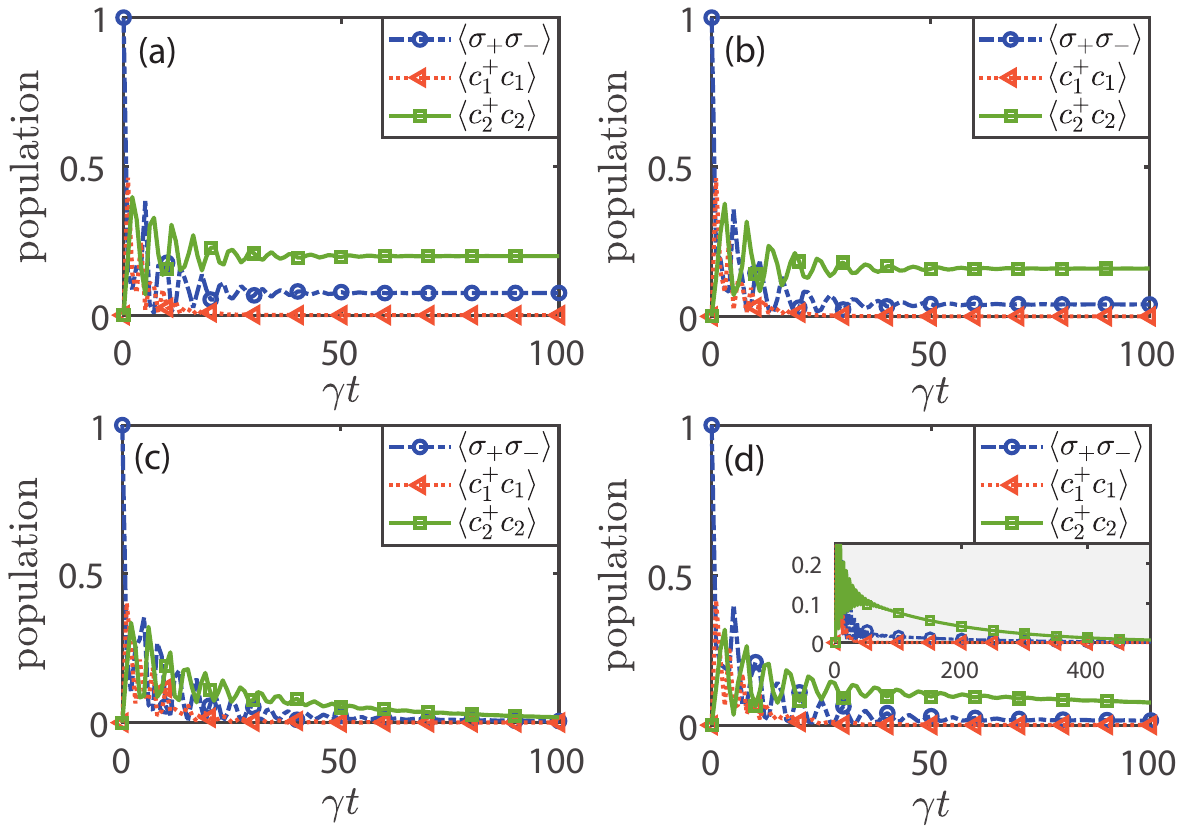}}
   \caption{Non-Markovian time-dependent evolution of the population in different parameters obtained by numerically solving Eq.~(\ref{dotABCtNonMarkovian}). (a) and (b) satisfy the conditions of vacancy-like dressed bound state, while (c) and (d) do not. The parameters chosen are $g=\gamma$, $J=\gamma$, (a) $\theta=\pi/3$, $\Delta_{0c}=-J \cos \theta$; (b) $\theta=\pi/4$, $\Delta_{0c}=-J \cos \theta$; (c) $\theta=\pi/3$, $\Delta_{0c}=-J$; (d) $\theta=\pi/4$, $\Delta_{0c}=-J$.}\label{popualtionnonMarkovian}
 \end{figure}

The spontaneous emission spectrum $S_{NM}(\omega)$ as a function of $\omega$ with various detuning $\Delta_{0c}=\omega_0-\omega_c$ is shown by Fig.~\ref{nMfig3}, which is solved by Eq.~(\ref{SEspectrumNonMarkovian}). Figures~\ref{nMfig3}(a) and \ref{nMfig3}(c) correspond to $\theta=0$, while Figs.~\ref{nMfig3}(b) and \ref{nMfig3}(d) are for $\theta=\pi/3$.
In Fig.~\ref{nMfig3}(a) with $\theta=0$, the $c_2$ mode decouples from the rest, remaining in a vacuum state. Consequently, there are no vacuum-dressed bound states, and only two peaks appear in the emission spectrum as $\Delta_{0c}$ varies in Fig.~\ref{nMfig3}(a). In Fig.~\ref{nMfig3}(b), $\theta=\pi/3$ modifies the coupling conditions, enabling vacuum-dressed bound states for detunings satisfying $\Delta_{0c}=-J \cos \theta=-0.5\gamma$ as indicated by the red line. While the green and blue lines exhibit sharp resonant peaks around $\omega=-\gamma$, the central peak associated with the red line is conspicuously absent. This absence signifies a disruption in the formation of vacuum-dressed bound states. Figure~\ref{nMfig3}(c) considers the behavior of the system in the Markovian limit for $\lambda=50\gamma$, where memory effects are minimal. In this regime, the previously observed peaks are substantially reduced, which is consistent with the expected Markovian dynamics.

In Fig.~\ref{popualtionnonMarkovian}, we plot non-Markovian evolution of the population in different parameters. We take $\theta=\pi/3$ in Fig.~\ref{popualtionnonMarkovian}(a) and (c), and $\theta=\pi/4$ in Fig.~\ref{popualtionnonMarkovian}(b) and (d). When the conditions $\theta=\pi/3$ and $\Delta_{0c}=-J \cos \theta$ are satisfied, the steady-state population of $\langle \sigma_+ \sigma_- \rangle$ (blue-circle line) and $\langle c_2^\dag c_2 \rangle$ (green-square line) remain finite, while the population of $\langle c_1^\dag c_1 \rangle$ (red-triangle line) is depleted and finally approachs zero. In addition, similar trends are observed for $\theta=\pi/4$ as shown in Fig.~\ref{popualtionnonMarkovian}(a) and Fig.~\ref{popualtionnonMarkovian}(b). However, the population of the atom exhibits an oscillating damping process in Fig.~\ref{popualtionnonMarkovian}(c) and (d) due to unmet vacancy-like dressed bound state condition $\Delta_{0c}=-J$, which is similar to Markovian environment.

%\begin{figure}[htbp]
%   \centerline{
%   \includegraphics[width=9cm, height=4.2cm, clip]{nonMarkovian_fig1.eps}}
%   \caption{Non-Markovian evolution of the population in different parameters obtained by numerically solving Eq.~(x). The parameters chosen are $\theta=\pi/3$, $\Delta_{0c}=-J\cos \theta$, (a) $J=\lambda=\gamma$; (b) $g=\lambda=\gamma$; (c) $J=g=\gamma$.}\label{popualtion1nonMarkovian}
% \end{figure}

%figure 10
 \begin{figure}[t]
   \centerline{
   \includegraphics[width=8.4cm, height=4.6cm, clip]{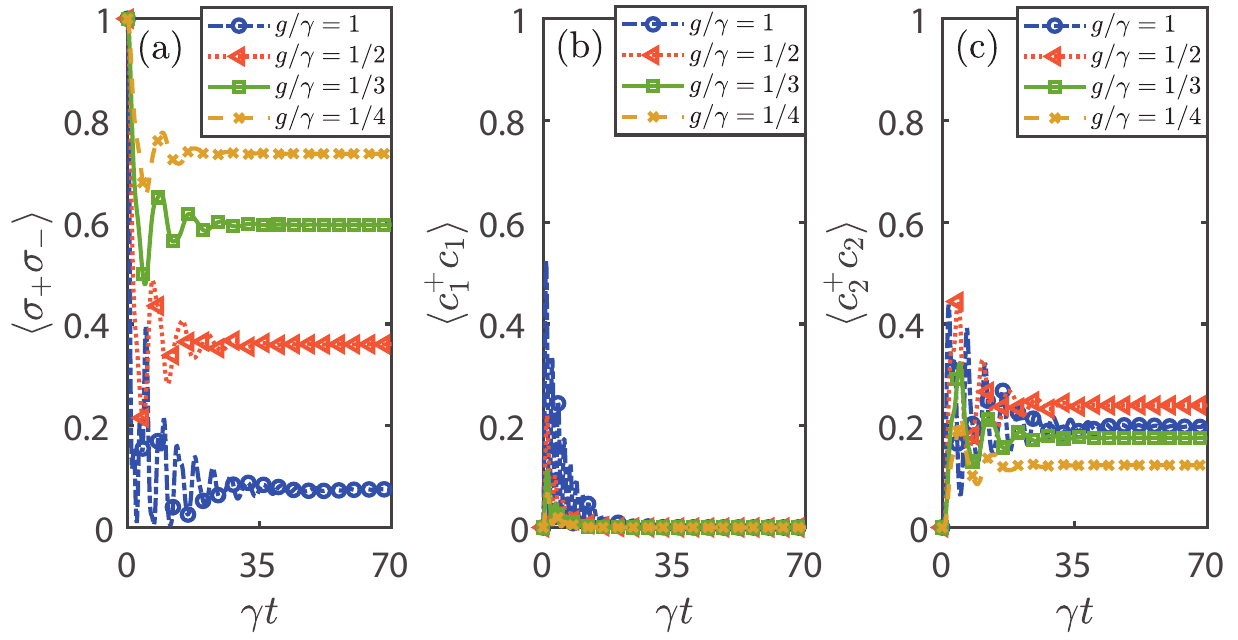}}
   \caption{Non-Markovian time-dependent evolution of the population in different atom-cavity coupling strength $g$ obtained by numerically solving Eq.~(\ref{dotABCtNonMarkovian}). The parameters chosen are $\theta=\pi/3$, $\Delta_{0c}=-J\cos \theta$, and $J=\lambda=\gamma$.}\label{gnonMarkovian}
 \end{figure}

The variation of $\langle \sigma_+ \sigma_- \rangle$, $\langle c_1^\dag c_1 \rangle$ and $\langle c_2^\dag c_2 \rangle$ with respect to the time $t$ for different coupling strength $g$ is plotted in Fig.~\ref{gnonMarkovian}.
%The different lines correspond to different values $g/\gamma  = 1$ (blue-circleline), $g/\gamma  = 1/2$ (red-triangle line), $g/\gamma  = 1/3$ (green-squareline), and $g/\gamma  = 1/4$ (yellow-fork line).
The steady state atom population decreases with the increase of $g$ in Fig.~\ref{gnonMarkovian}(a) with the other parameters fixed, while the steady state population of mode ${{{\hat c}_1}}$ remains zero as $g$ varies in Fig.~\ref{gnonMarkovian}(b).
In Fig.~\ref{gnonMarkovian}(c), the steady state atom population of mode ${{{\hat c}_2}}$ increases with the increase of $g$.
Indeed, as the coupling strength $g$ between the atoms and the optical cavity increases, a greater number of photons are coupled into the cavity. This results in a reduction of the atomic population in the steady state and an enhancement of the population in the ${{{\hat c}_2}}$ configuration.

%figure 11
 \begin{figure}[t]
   \centerline{
   \includegraphics[width=8.4cm, height=4.6cm, clip]{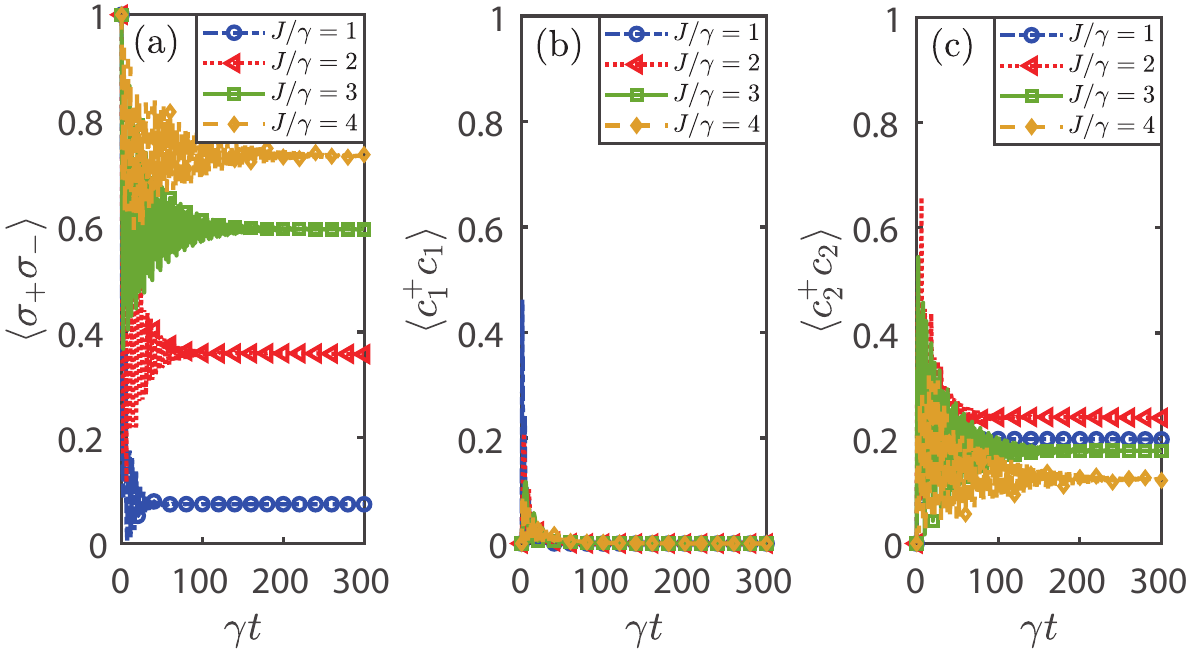}}
   \caption{Non-Markovian time-dependent evolution of the population in various coupling strength $J$ between two modes obtained by numerically solving Eq.~(\ref{dotABCtNonMarkovian}). The parameters chosen are $\theta=\pi/3$, $\Delta_{0c}=-J\cos \theta$, $g=\lambda=\gamma$.}\label{JnonMarkovian}
 \end{figure}

\begin{figure}[t]
   \centerline{
   \includegraphics[width=8.4cm, height=4.6cm, clip]{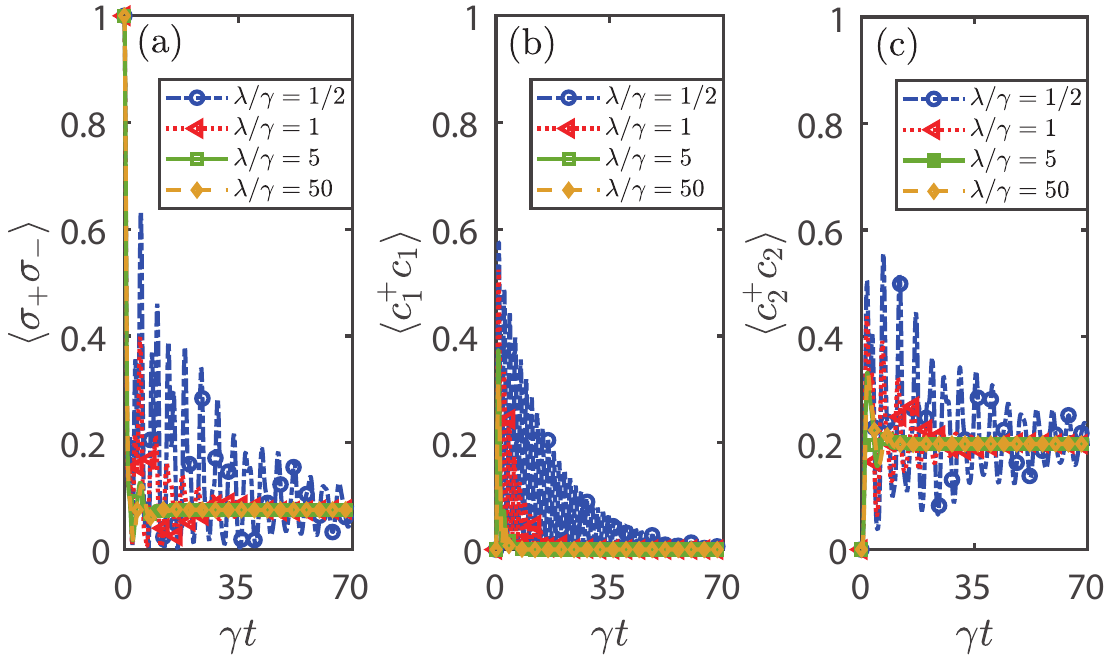}}
   \caption{Non-Markovian  time-dependent evolution of the population in different environmental spectrum width $\lambda$ obtained by numerically solving Eq.~(\ref{dotABCtNonMarkovian}). The parameters chosen are $\theta=\pi/3$, $\Delta_{0c}=-J\cos \theta$, and $J=g=\gamma$.}\label{lambdanonMarkovian}
 \end{figure}

Next, we consider the case of $g=\gamma$, where the probability amplitudes $\langle \sigma_+ \sigma_- \rangle$, $\langle c_1^\dag c_1 \rangle$ and $\langle c_2^\dag c_2 \rangle$ as functions of $t$ for different values of $J$ are plotted in Fig.~(\ref{JnonMarkovian}). In contrast to the scenario where $g$ alters, the amplitude of the atomic steady state population $\langle \sigma_+ \sigma_- \rangle$ increases with the increase in $J$ in Fig.~\ref{JnonMarkovian}(a). This behavior is straightforward to understand. An increase in the coupling strength $J$ between CW mode and CCW mode within the optical cavity implies according to the vacancy-like bound state condition ${\omega _0} - {\omega _c} =  - J\cos \theta $ that the frequency discrepancy between the optical cavity $\omega_c$ and the atom $\omega_0$ will also widen. This results in a reduced probability of photon entrapment by the cavity, which in turn causes an augmentation in the atomic steady-state population and a concomitant diminution in the population of the $c_2$ state. As illustrated in Fig.~\ref{lambdanonMarkovian}, there is a progressive decrease in the relaxation time with the increase of the spectral width $\lambda$ varying from $\gamma/2$ to $50\gamma$. However, this increase in the spectral width does not influence the magnitude of the steady-state probability amplitude.

%figure 12

%\subsection{Friedrich-Wintgen dressed bound state}
Friedrich and Wintgen demonstrated that bound states in the continuum can occur due to the interference of different resonances \cite{Friedrich3231}. If two resonances have a degeneracy point when we tune some continuous parameter, the interference can cause an avoided level crossing of the frequencies, where a bound state in the continuum with vanishing resonance width may be formed at some specific value of the continuous parameter.

The effective Hamiltonian for an open photonic system has been studied previously, which is non-Hermitian and can be written as follows:
\begin{equation}
\begin{aligned}
   Q_{NMc}={H}_{+}-i {H}_{-},
\label{nonFWDBSdecompose}
\end{aligned}
\end{equation}
with
\begin{equation}
\begin{aligned}
   {H}_{+} &=\left( {\begin{array}{*{20}{c}}
{{\omega _0} - {\omega _c}}&g&g&0\\
g&0&{J{e^{i\theta }}}&{ - i(\frac{{\Gamma \lambda }}{4} + \frac{1}{2})}\\
g&{J{e^{ - i\theta }}}&0&{ - i(\frac{{\Gamma \lambda }}{4} + \frac{1}{2})}\\
0&{i(\frac{{\Gamma \lambda }}{4} + \frac{1}{2})}&{i(\frac{{\Gamma \lambda }}{4} + \frac{1}{2})}&0
\end{array}} \right),\\
{H}_{-} &=\left( {\begin{array}{*{20}{c}}
0&0&0&0\\
0&0&0&{(\frac{1}{2} - \frac{{\Gamma \lambda }}{4})}\\
0&0&0&{(\frac{1}{2} - \frac{{\Gamma \lambda }}{4})}\\
0&{(\frac{1}{2} - \frac{{\Gamma \lambda }}{4})}&{(\frac{1}{2} - \frac{{\Gamma \lambda }}{4})}&{\lambda}
\end{array}} \right),
\label{nonFWDBSreimag}
\end{aligned}
\end{equation}
where ${H}_{+}$ is a Hermitian operator giving rise to discrete and real eigenvalues for the bound states, while ${H}_{-}$ denotes an anti-Hermitian part governing the imaginary part of the eigenenergies. When these eigenstates couple to some open channels characterized by the coupling matrix $D$, the energy will leak out and the eigenvalues of energy are no longer purely real. However, we note that the rank of matrix $r({H}_{-})=2$ is less than the dimension of $H_{-}$, which leads to there is no matrix $D$ that makes $D^\dagger D=H_{-}$. In addition, there is no non-zero vector $|\psi_0 \rangle$ satisfying $H_{-} |\psi_0 \rangle=0$. Therefore, there is no vector $|\psi_0 \rangle$ having a purely real eigenenergy and satisfying $Q_{NMc} |\psi_0 \rangle = \omega_0 |\psi_0 \rangle$, namely, there is no Friedrich-Wintgen dressed bound state in the non-Markovian regimes.

\subsection{The case of non-zero accumulated phase factor}

Previously, we analyzed a simple situation, and now we will return to a more general situation given by Eq.~(\ref{dotABCtjfNonMarkovian}). Without loss of generality, we transform back to Schr\"{o}dinger picture, which leads to Eq.~(\ref{dotABCtjfNonMarkovian}) becoming
\begin{equation}
\begin{aligned}
i\dot A(t) =& {\omega _0} A(t) + gB(t) + gC(t),\\
i\dot B(t) =& gA(t) + {\omega _c} B(t) + JC(t) - i\int_0^t {d\tau } {f_1}(t - \tau )B(\tau )\\
& - i\int_0^t {d\tau } {f_2}(t - \tau )C(\tau ),\\
i\dot C(t) =& gA(t) + JB(t) + {\omega _c} C(t) - i\int_0^t {d\tau } {f_3}(t - \tau )B(\tau )\\
& - i\int_0^t {d\tau } {f_1}(t - \tau )C(\tau ).
\label{dotdtNonMarkovianne2}
\end{aligned}
\end{equation}
We note that whether to perform a rotating frame transformation does not affect the square of the probability amplitude, as seen by Appendix \ref{APPC}. To proceed, we perform a Laplace transform to Eq.~(\ref{dotdtNonMarkovianne2}) and obtain
\begin{equation}
\begin{aligned}
-i &= ({\omega _0}-is) \widetilde{A}(s) + g \widetilde{B}(s) + g \widetilde{C}(s),\\
0 &= g \widetilde{A}(s) + ({\omega _c}-is- i \widetilde{f_1}(s)) \widetilde{B}(s) + (J - i \widetilde{f_2}(s)) \widetilde{C}(s),\\
0 &= g \widetilde{A}(s) + (J- i \widetilde{f_3}(s)) \widetilde{B}(s) + ({\omega _c}-is- i \widetilde{f_1}(s)) \widetilde{C}(s),
\label{NonMarkovianL}
\end{aligned}
\end{equation}
with the initial conditions $A(0)=1$ and $B(0)=C(0)=0$, where $\widetilde{f_1}(s) = \int {d\omega \frac{{{{| {\gamma_k} |}^2}}}{{s + i\omega }}} $, $\widetilde{f_2}(s) = \int {d\omega \frac{{{{| {\gamma_k} |}^2}e^{i k x_1}}}{{s + i\omega }}} $ and $\widetilde{f_3}(s) = \int {d\omega \frac{{{{| {\gamma_k} |}^2}e^{-i k x_1}}}{{s + i\omega }}} $. Solving Eq.~(\ref{NonMarkovianL}) obtains
\begin{equation}
\begin{aligned}
\widetilde{A}(s) &= \frac{\eta_1}{i g^2 \eta_2 + (s+i\omega_0)\eta_1},\\
\widetilde{B}(s) &= \frac{(is + i \widetilde{f_1}(s) -i \widetilde{f_2}(s) - \omega_c + J)g}{i g^2 \eta_2 + (s+i\omega_0)\eta_1},\\
\widetilde{C}(s) &= \frac{(is + i \widetilde{f_1}(s) -i \widetilde{f_3}(s) - \omega_c + J)g}{i g^2 \eta_2 + (s+i\omega_0)\eta_1},
\label{NonMarkovianS}
\end{aligned}
\end{equation}
with $\eta_1 = [is +i \widetilde{f_1}(s) -\omega_c]^2 - [J-i \widetilde{f_2}(s)][J-i \widetilde{f_3}(s)]$ and $\eta_2 = 2[is +i \widetilde{f_1}(s) -\omega_c] + [J-i \widetilde{f_2}(s)]+[J-i \widetilde{f_3}(s)]$. According to the Cauchy residue theorem, the inverse Laplace transform can be performed by finding all the poles of $\widetilde{A}(s)$, $\widetilde{B}(s)$ and $\widetilde{C}(s)$. We now consider a special case where there is a pole on the imaginary axis, i.e., purely imaginary axis $s = - i E_{bs}$ ($E_{bs}$ is a real number) in which poles equation,
\begin{equation}
i g^2 \eta_2 + (s+i\omega_0)\eta_1 = 0,
\label{NonMarkovianL1}
\end{equation}
leads to the identity,
\begin{equation}
\begin{aligned}
&2 g^2 [E_{bs}-\omega_c-A_1(E_{bs})+J+\frac{A_2(E_{bs})+A_3(E_{bs})}{2}] \\
= &(E_{bs} - \omega_0)\{[E_{bs}-\omega_c-A_1(E_{bs})]^2-[J+A_2(E_{bs})]\\
& \times [J+A_3(E_{bs})]\},
\label{NonMarkovianL2}
\end{aligned}
\end{equation}
where $A_1(E_{bs})=-i \widetilde{f_1}(E_{bs}) = \int {d\omega \frac{{{{| {\gamma_k} |}^2}}}{{E_{bs} - \omega }}}$, $A_2(E_{bs})=-i \widetilde{f_2}(E_{bs}) = \int {d\omega \frac{{{{| {\gamma_k} |}^2}e^{i k x_1}}}{{E_{bs} - \omega }}}$ and $A_3(E_{bs})=-i \widetilde{f_3}(E_{bs}) = \int {d\omega \frac{{{{| {\gamma_k} |}^2}e^{-i k x_1}}}{{E_{bs} - \omega }}}$.
For the sake of discussion about the critical equation of the bound states below, we reorganize Eq.~(\ref{NonMarkovianL2}) and define
\begin{align}
&X(E_{bs})\nonumber\\
= &(E_{bs} - \omega_0)\{[E_{bs}-\omega_c-A_1(E_{bs})]^2-[J+A_2(E_{bs})]\nonumber\\
& \times [J+A_3(E_{bs})]\}- 2 g^2 J + 2 g^2 \omega_c + 2 g^2 A_1(E_{bs})\nonumber\\
& -g^2 (A_2(E_{bs})+A_3(E_{bs})) .
\label{NonMarkovianL3}
\end{align}
Since $X(E_{bs})$ is a monotonically decreasing function when $E_{bs} < 0$, Eq.~(\ref{NonMarkovianL2}) has one discrete root if $X(0) < 0$. It has an infinite number of roots in the regime $E_{bs} > 0$, which form a continuous energy band. We name this discrete eigenstate with eigenenergy $E_{bs} < 0$ the bound state. Its formation would have profound consequences on the decoherence dynamics.

Note that the roots of Eq.~(\ref{NonMarkovianL2}) are just the eigenenergies in the single-excitation subspace of the whole system consisting of the atom, the optical cavity and semi-infinite waveguide. To see this, we expand the eigenstates as $| \Phi \rangle  = \mathfrak{A}| {{{e,0,0,}}{{{0}}}} \rangle  + \mathfrak{B}| {{{g,1,0,}}{{{0}}}} \rangle + \mathfrak{C}| {{{g,0,1,}}{{{0}}}} \rangle  + \int {d\omega \mathfrak{D}_k\hat{b}_k^\dag  | {{{g,0,0,}}{{{0}}}} \rangle }$. Then from the eigenequation
\begin{equation}
\hat{H}_{NM} | \Phi \rangle = E | \Phi \rangle,
\label{NonMarkovianExpand}
\end{equation}
with the Hamiltonian given by Eq.~(\ref{HamiltonianNonMarkovian}), we obtain
\begin{subequations}
\begin{align}
E \mathfrak{A} &= \omega_0 \mathfrak{A} + g \mathfrak{B} + g \mathfrak{C}, \label{NonMarkovianS1}\\
E \mathfrak{B} &= g \mathfrak{A} + \omega_c \mathfrak{B} + J \mathfrak{C} - i \int {d\omega \gamma^*(\omega) e^{i k x_{cw}} \mathfrak{D}_k}, \label{NonMarkovianS2}\\
E \mathfrak{C} &= g \mathfrak{A} + J \mathfrak{B} + \omega_c \mathfrak{C} - i \int {d\omega \gamma^*(\omega) e^{i k x_{ccw}} \mathfrak{D}_k}, \label{NonMarkovianS3}\\
E \mathfrak{D}_k &= \omega \mathfrak{D}_k + i \gamma(\omega) e^{-i k x_{cw}} \mathfrak{B} + i \gamma(\omega) e^{-i k x_{ccw}} \mathfrak{C}. \label{NonMarkovianS4}
\end{align}
\end{subequations}
By substituting Eq.~(\ref{NonMarkovianS4}) into Eqs.~(\ref{NonMarkovianS2}) and~(\ref{NonMarkovianS3}), the following linear system of equations can be obtained
\begin{equation}
\begin{aligned}
0 = \ &(\omega_0 - E) \mathfrak{A} + g \mathfrak{B} + g \mathfrak{C}, \\
0 = \ &g \mathfrak{A} + (\omega_c - E + \int {d\omega \frac{{{{| {\gamma_k} |}^2}}}{{E - \omega }}}) \mathfrak{B} \\
   \ &+ (J + \int {d\omega \frac{{{{| {\gamma_k} |}^2}e^{i k x_1}}}{{E - \omega }}}) \mathfrak{C},\\
0 = \ &g \mathfrak{A} + (J + \int {d\omega \frac{{{{| {\gamma_k} |}^2}e^{-i k x_1}}}{{E - \omega }}}) \mathfrak{B} \\
   \ &+ (\omega_c - E + \int {d\omega \frac{{{{| {\gamma_k} |}^2}}}{{E - \omega }}}) \mathfrak{C}. \label{NonMarkovianS5}
\end{aligned}
\end{equation}
The above linear equation system has non-zero solutions if and only if the coefficient determinant is zero, which also leads to Eq.~(\ref{NonMarkovianL2}). In this regime with ${|\mathfrak{A}|}^2 + {|\mathfrak{B}|}^2 + {|\mathfrak{C}|}^2 + \int {d\omega {|\mathfrak{D}_k|}2} = 1$, we obtain
\begin{align}
\mathfrak{A}_{bs}= \ &\{1 +\frac{(E-\omega_0)(B_2 A_3 + B_3 A_2)}{\eta_1} -\frac{{2(1 + B_1)\eta_3}}{\eta_1}\nonumber\\
               &  +\frac{{[g^2 + (E - \omega_0)J]}{(B_2+B_3)}}{\eta_1} \}^{-1/2},
\label{Abs}
\end{align}
%\begin{equation}
%\begin{aligned}
%\mathfrak{A}_{bs}=&\{1+\frac{{(1 + B_1)[\eta_3^2 + |{g^2} + (A_2 + J)(E - {\omega _0})|^2]}}{{{g^2}| - A_1 + A_2 + E + J - {\omega _c}|^2}} \\
%               &-\frac{\eta_3(E-\omega_0)[B_2 A_3 + B_3 A_2+(B_2+B_3)J] }{{g^2}| - A_1 + A_2 + E + J - {\omega _c}|^2}\\
%               &-\frac{{\eta_3}{(B_2+B_3)}}{| - A_1 + A_2 + E + J - {\omega _c}|^2}\}^{-1/2},
%\label{Abs}
%\end{aligned}
%\end{equation}
where $\eta_3 = {g^2}+ (E - {\omega _0})(A_1 - E + {\omega _c})$, $B_1 = \int {d\omega \frac{{{{| {\gamma_k} |}^2}}}{(E - \omega )^2}}$, $B_2 = \int {d\omega \frac{{{{| {\gamma_k} |}^2}e^{i k x_1}}}{(E - \omega )^2}}$, and $B_3 = \int {d\omega \frac{{{{| {\gamma_k} |}^2}e^{-i k x_1}}}{(E - \omega )^2}}$. It is interesting to see that the roots of Eq.~(\ref{NonMarkovianL2}) are just the eigenenergies in the single-excitation subspace. It is understandable from the fact that the decoherence of the atom and optical cavity induced by the vacuum environment is governed by the single-excitation process of the whole system.

To see this, we perform the inverse Laplace transform to $\widetilde{A}(s)$, and obtain
\begin{equation}
A(t)=|\mathfrak{A}_{bs}|^2 e^{-i E_{bs}t} + \int_{i \varsigma + 0}^{i \varsigma + \infty} {{\frac{d\omega}{2\pi}}\widetilde{A}(-i \omega)e^{-i\omega t}},
\label{At}
\end{equation}
where $\mathfrak{A}_{bs}$ is given by Eq.~(\ref{Abs}) and the second term contains contributions from the continuous energy band. Oscillating with time in continuously changing frequencies, the second term in Eq.~(\ref{At}) decays and tends to zero due to out-of-phase interference. Therefore, if the bound state is absent, then $\mathop {\lim }\limits_{t \to \infty} {A(t)}=0$ characterizes a complete decoherence, whereas if the bound state is formed, then $\mathop {\lim }\limits_{t \to \infty} {A(t)}=|\mathfrak{A}_{bs}|^2 e^{-i E_{bs}t}$ implies dissipation suppression.

To make this result clear, we recall that, according to the Schr\"{o}dinger equation $i|\dot{\psi}(t)\rangle=\hat{H}|\psi(t)\rangle$, the time evolution of the whole state satisfies $|\psi(t)\rangle = e^{-i\hat{H}t} |e,0,0,0\rangle$. Inserting completeness relations $|\Phi_{bs}\rangle \langle\Phi_{bs}| + \int {|\Phi_{c}\rangle \langle\Phi_{c}| dE_c} = I$ ($I$ is an identity matrix) into it, we obtain
\begin{equation}
|\psi(t)\rangle = \mathfrak{A}_{bs} e^{-i E_{bs}t} |\Phi_{bs}\rangle + |\psi_c (t)\rangle,
\label{psitbsc}
\end{equation}
where $|\Phi_{bs}\rangle$ denotes the bound state with energy $E_{bs}$ given by Eq.~(\ref{NonMarkovianL2}). $|\psi_{c}\rangle$ is a superposition of the continuous spectrum eigenfunctions of the Hamiltonian,
\begin{equation}
|\psi_c (t)\rangle = \int {dE_c e^{-i E_c t} \mathfrak{A}_c^* |\psi_c (E_c)\rangle},
\label{psic}
\end{equation}
with the continuous-spectrum eigenfunctions
\begin{equation}
\begin{aligned}
|\Phi_c (E_c)\rangle =& \mathfrak{A}_c |e,0,0,0\rangle + \mathfrak{B}_c |g,1,0,0\rangle \\
                      &+ \mathfrak{C}_c |g,0,1,0\rangle + \int {d E_c \mathfrak{D}_{k,c} |g,0,0,1_k\rangle},
\label{Phic}
\end{aligned}
\end{equation}
where $E_c$ denotes the continuous-spectrum eigenenergy in the regime of $E_c > 0$, which is obtained by the diagonalization of
\begin{widetext}
\begin{equation}
\begin{aligned}
\hat{H}_{se}=\left( {\begin{array}{*{20}{c}}
{{\omega _0}}&g&g&0&0&0\\
g&{{\omega _c}}&J&{i{\gamma _1}{e^{i{k_1}{x_{cw}}}}}&{\cdots}&{i{\gamma _N}{e^{i{k_N}{x_{cw}}}}}\\
g&J&{{\omega _c}}&{i{\gamma _1}{e^{i{k_1}{x_{ccw}}}}}&{\cdots}&{i{\gamma _N}{e^{i{k_N}{x_{ccw}}}}}\\
0&{i{\gamma _1}{e^{-i{k_1}{x_{cw}}}}}&{i{\gamma _1}{e^{-i{k_1}{x_{ccw}}}}}&{{\omega _1}}&0&0\\
0&{\cdots}&{\cdots}&0&{\cdots}&0\\
0&{i{\gamma _N}{e^{-i{k_N}{x_{cw}}}}}&{i{\gamma _N}{e^{-i{k_N}{x_{ccw}}}}}&0&0&{{\omega _N}}
\end{array}} \right),
\label{Hcc}
\end{aligned}
\end{equation}
\end{widetext}
where $N$ denotes total mode numbers of the structured environment. $\mathfrak{A}_c$. $\mathfrak{B}_c$, $\mathfrak{C}_c$ and $\mathfrak{D}_{k,c}$ are solved in the regime of $E_c > 0$ by Eq.~(\ref{NonMarkovianS5}). The probability amplitude on the upper state from Eq.~(\ref{psitbsc}) can be written as
\begin{equation}
|\mathfrak{A}_{bs}|^2 e^{-i E_{bs}t} + \int {dE_c e^{-i E_c t} {|\mathfrak{A}_c|}^2 }.
\label{psitbsc2}
\end{equation}
We show that the second term of Eq.~(\ref{psitbsc2}) corresponds to the second term of Eq.~(\ref{At}), which tends to zero in the longtime limit $t \to \infty$ according to the Lebesgue-Riemann lemma \cite{Bochner1949}. This leads to the probability amplitude of the upper state to approach $|\mathfrak{A}_{bs}|^2 e^{-i E_{bs}t}$. Note that this feature has been illustrated in the literature \cite{Liu052139,Tong174301} to describe the incomplete decay of an atom in photonic bandgap media.

\begin{figure}[t]
   \centerline{
   \includegraphics[width=8.6cm,clip]{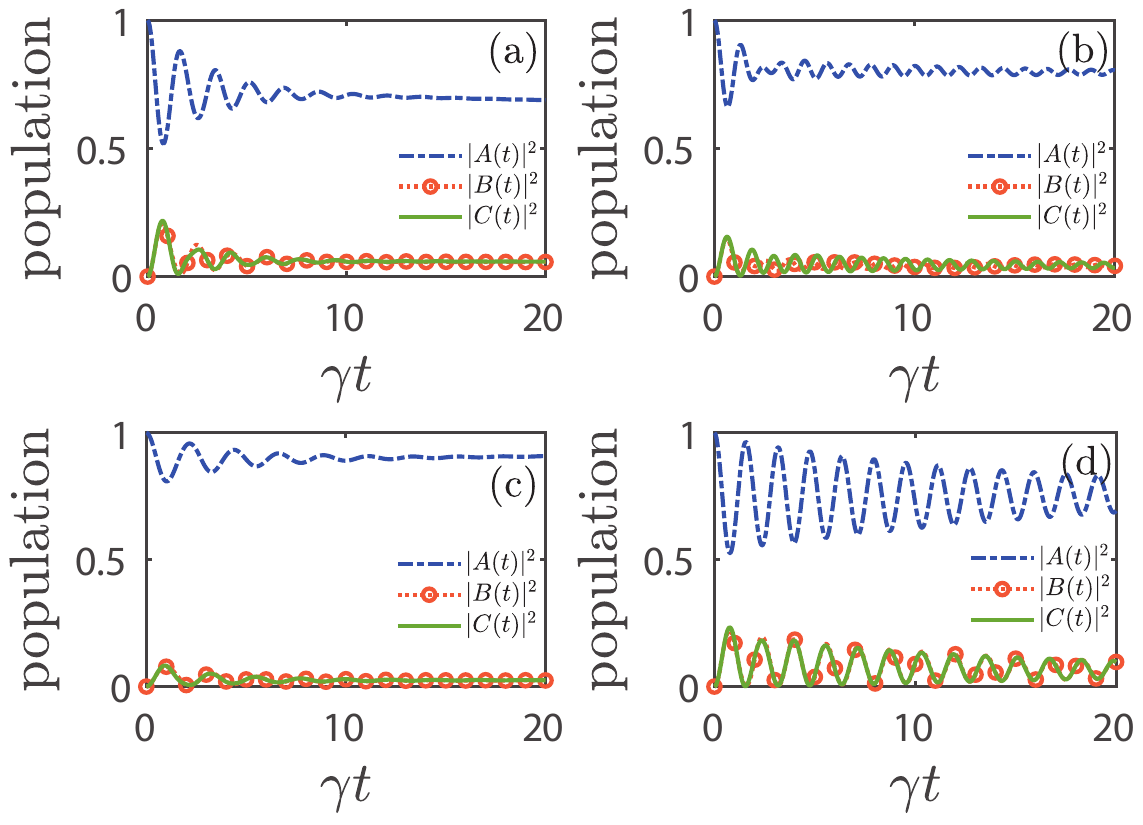}}
   \caption{Non-Markovian  time-dependent evolution of the population in different parameters obtained by solving Eq.~(\ref{dotABCtjfNonMarkovian}). The parameters of (a) chosen are $\Phi = 5$, $\lambda = 2\gamma$, $x_1/v = \gamma$, $g = \gamma$, $J = 2\gamma$, $\Delta_{0c} = 5\gamma$. The parameter of (b) chosen is $J=\gamma$, where the other parameters are the same as in (a). (c) $g=0.5\gamma$, where the other parameters are the same as in (a). (d) $\lambda=0.5\gamma$ with the other parameters being the same as in (a). }\label{popualtionnonMarkovian1}
 \end{figure}

Figure~\ref{popualtionnonMarkovian1} shows non-Markovian evolution of the population for different parameters obtained by solving Eq.~(\ref{dotABCtjfNonMarkovian}) under the condition (\ref{NonMarkovianL2}). The population stabilizes over time as shown in Fig.~\ref{popualtionnonMarkovian1}(a). If we change $J=1\gamma$, the steady state atom population increases with the increase of $J$, which is different from vacancy-like bound state in the case of zero accumulation phase factor. This phenomenon arises due to the absence of the vacancy-like bound state condition as described by Eq.~(\ref{nonMarkovianVLcondition}). In this context, fixing $\Delta_{0c}$, the coupling between the two optical modes is reduced as $J$ decreases, whereas the coupling between the optical cavity and the non-Markovian environment remains unchanged. Consequently, this leads to an increase in the atomic population, while the population of the two optical cavity modes diminishes. Taking $g=0.5\gamma$ in Fig.~\ref{popualtionnonMarkovian1}(c), the steady state atom population decreases with the increase of $g$, which is consistent with the case of the vacancy-like bound state. However, reducing $\lambda$ to $\lambda=0.5\gamma$ significantly increases the relaxation time as shown in Fig.~\ref{popualtionnonMarkovian1}(d).

%figure 13

\section{conclusion}

In conclusion, we have demonstrated and unveiled the origin of DBSs in a prototypical microring resonator, which are divided into two types, the vacancy-like and Friedrich-Wintgen-type bound states. DBSs studied in this work exist in the single-photon manifold, while the principles can be applied to higher-excitation manifolds for exploring multiphoton DBSs. Moreover, we extend the study of vacancy-like dressed bound state and Friedrich-Wintgen dressed bound state from the Markovian to the non-Markovian bath, which consists of a collection of infinite oscillators (bosonic photonic modes). The peak value and relaxation time of the vacancy-like dressed bound state in the non-Markovian regime with zero accumulated phase factor are higher than those in the Markovian regime, while there is no Friedrich-Wintgen dressed bound state for the non-Markovian case. More generally, we verify the correspondence between the energy spectrum and bound state conditions for the case of non-zero accumulated phase factor, and study the influence of various parameters on the non-Markovian bound states.

Our results are experimentally observable in the WGM optical cavity platform, where the bound states have been observed already. Note that current studies might be extendable to non-Hermitian non-Markovian system (e.g., WGM microdisk perturbed by two or more nanoparticles instead of one nanoparticle \cite{Xu023803,Wiersig063828}), and hence it is interesting for the community of quantum physics.

\section*{ACKNOWLEDGMENTS}
H. Z. S. acknowledges National Natural Science Foundation of China under Grants No.~12274064 and Scientific Research Project for Department of Education of Jilin Province under Grant No.~JJKH20241410KJ. C. S. acknowledges financial support from the China Scholarship Council, the Japanese Government (Monbukagakusho-MEXT) Scholarship (Grant No.~211501), the RIKEN Junior Research Associate Program, and the Hakubi Projects of RIKEN. Y. H. Z. acknowledges the National Natural Science Foundation of China (NSFC)(Grants Nos.~12374333).

\section*{}
\appendix
\begin{widetext}
\section{\label{APPA} The derivation details of Eq.~(\ref{dotABCt})}
Substituting Eq.~(\ref{expressstate}) into Schr\"{o}dinger equation $i|\dot{\psi}(t)\rangle=\hat{H}|\psi(t)\rangle$, we obtain
%\begin{small}
\begin{subequations}
\begin{align}
i\dot A(t) &= ({\omega _0} - {\omega _c})A(t) + gB(t) + gC(t),\label{APPAdotA}\\
i\dot B(t) &= gA(t) + J C(t) - i \int {d\omega \sqrt{\frac{\gamma}{2\pi}}e^{-i(\omega-\omega_c)t}e^{ikx_{cw}}D_k(t)},\label{APPAdotB}\\
i\dot C(t) &= gA(t) + JB(t) - i \int {d\omega \sqrt{\frac{\gamma}{2\pi}}e^{-i(\omega-\omega_c)t}e^{ikx_{ccw}}D_k(t)},\label{APPAdotC}\\
i\dot D_k(t) &= i\sqrt{\frac{\gamma}{2\pi}}e^{i(\omega-\omega_c)t}e^{-ikx_{cw}}B(t) +  i\sqrt{\frac{\gamma}{2\pi}}e^{i(\omega-\omega_c)t}e^{-ikx_{ccw}}C(t).\label{APPAdotD}
\end{align}
\end{subequations}
%\end{small}
Eq.~(\ref{APPAdotD}) can be formally integrated to obtain the equation of motion of $D_k(t)$

\begin{equation}
\begin{aligned}
D_k(t) = \sqrt{\frac{\gamma}{2\pi}}[\int{d \tau e^{i(\omega-\omega_c)t} e^{-ikx_{cw}} B(t)} +  \int{d\tau e^{i(\omega-\omega_c)t} e^{-ikx_{ccw}} C(t)}], \label{APPAD}
\end{aligned}
\end{equation}
where we have taken $D_k(0) = 0$ since the waveguide is initially prepared in the vacuum state. Substituting $D_k(t)$ into the last term of Eq.~(\ref{APPAdotB}), we have
\begin{equation}
\begin{aligned}
 &- i \int {d\omega \sqrt{\frac{\gamma}{2\pi}}e^{-i(\omega-\omega_c)t}e^{ikx_{cw}}D_k(t)}\\
 =&-i\frac{\gamma}{2\pi}[\int{d \tau} \int{d\omega e^{i(\omega-\omega_c)(\tau-t)} B(\tau)} +  \int{d \tau} \int{d\omega e^{i(\omega-\omega_c)(\tau-t)} e^{ik(x_{cw}-x_{ccw})} C(\tau)}]\\
 = & -i\frac{\gamma}{2\pi}[\int_0^t {d \tau} 2\pi \delta(\tau-t) B(\tau) +  \int{d \tau} 2\pi \delta(\tau-t+x_1/v) e^{i\phi} C(\tau)]\\
 =& -i \frac{\gamma}{2}B(t) -i\gamma e^{i\phi} C(t), \label{APPAB1}
\end{aligned}
\end{equation}
where $x_1=x_{cw}-x_{ccw}>0$ and $\phi=\omega_c x_1/v$. We have applied the Markovian approximation by assuming the time delay $x_1/v$ between the CCW mode and the mirrored CW mode can be neglected. Therefore, the equation of motion of $B(t)$ can be obtained
\begin{equation}
\begin{aligned}
i\dot B(t) = gA(t) - i\frac{\gamma }{2}B(t) + (J - i\gamma {e^{i\phi }})C(t). \label{APPAB}
\end{aligned}
\end{equation}
Similarly, the differential equation for $C(t)$ can be derived as follows
\begin{equation}
\begin{aligned}
i\dot C(t) = gA(t) + JB(t) - i\frac{\gamma }{2}C(t). \label{APPAC}
\end{aligned}
\end{equation}
Therefore, Eq.~(\ref{dotABCt}) can be obtained by integrating Eqs.~(\ref{APPAdotA}), (\ref{APPAB}) and (\ref{APPAC}).

\section{\label{APPB} The derivation details of Eq.~(\ref{dotABCtNonMarkovian})}

Considering the case of zero accumulated phase factor, i.e., $x_{cw}=x_{ccw}$, Eq,~(\ref{dotABCtjfNonMarkovian}) can be rewritten as
\begin{equation}
\begin{aligned}
i\dot A(t) = &{\Delta _{0c}}A(t) + gB(t) + gC(t),\\
i\dot B(t) = &gA(t) + JC(t) - i\int_0^t {d\tau } {f_1}(t - \tau )B(\tau ) - i\int_0^t {d\tau {f_1}(t - \tau )C(\tau )},\\
i\dot C(t) = &gA(t) + JB(t) - i\int_0^t {d\tau } {f_1}(t - \tau )B(\tau )- i\int_0^t {d\tau {f_1}(t - \tau )C(\tau )} ,
\label{APPBdotABCtjfNonMarkovian}
\end{aligned}
\end{equation}
where ${f_1}(t - \tau ) = \frac{1}{2}{e^{ - \lambda \left| {t - \tau } \right|}}{\gamma}\lambda$. Defining $ W(t)  =\int_0^t {d\tau {f_1}(t - \tau )} B(\tau ) + \int_0^t {d\tau {f_1}(t - \tau )} C(\tau )$, we have
\begin{equation}
\begin{aligned}
\dot W(t)  = -\lambda W(t) + \frac{\gamma \lambda}{2} B(t) + \frac{\gamma \lambda}{2} C(t).
\label{APPBdotABCtjfNonMarkovian}
\end{aligned}
\end{equation}
Therefore, Eq.~(\ref{dotABCtNonMarkovian}) can be obtained by integrating Eqs.~(\ref{APPBdotABCtjfNonMarkovian}) and (\ref{APPBdotABCtjfNonMarkovian}).

\section{\label{APPC} Performing a rotating frame transformation preserves the square of the probability amplitude}

To clearly distinguish between the two cases, we denote the probability amplitudes without the rotating frame by $X(t)$, $Y(t)$ and $Z(t)$, while we use $A(t)$, $B(t)$ and $C(t)$ to denote the probability amplitudes within the rotating frame as shown by Eq.~(\ref{dotABCtjfNonMarkovian}). Eq.~(\ref{dotdtNonMarkovianne2}) can be rewritten as
\begin{equation}
\begin{aligned}
i\dot X(t) =& {\omega _0} X(t) + g Y(t) + g Z(t),\\
i\dot Y(t) =& g X(t) + {\omega _c} Y(t) + J Z(t) - i\int_0^t {d\tau } {\mathfrak{f}_1}(t - \tau ) Y(\tau ) - i\int_0^t {d\tau } {\mathfrak{f}_2}(t - \tau ) Z(\tau ),\\
i\dot Z(t) =& g X(t) + J Y(t) + {\omega _c} Z(t) - i\int_0^t {d\tau } {\mathfrak{f}_3}(t - \tau ) Y(\tau ) - i\int_0^t {d\tau } {\mathfrak{f}_1}(t - \tau ) Z(\tau ),
\label{APPCdotdtNonMarkovianne2}
\end{aligned}
\end{equation}
where $\mathfrak{f}_1(t-\tau) = \int {d\omega } {\left| {\gamma_k} \right|^2}{e^{ - i \omega (t - \tau )}}$, $\mathfrak{f}_2(t-\tau) = \int {d\omega } {\left| {\gamma_k} \right|^2}{e^{ - i \omega (t - \tau )}}{e^{ i k x_1}}$ and $\mathfrak{f}_3(t-\tau) = \int {d\omega} {\left| {\gamma_k} \right|^2}{e^{ - i \omega (t - \tau )}}{e^{- i k x_1}}$. Assuming $X(t)$, $Y(t)$ and $Z(t)$ are the solutions to Eq.~(\ref{APPCdotdtNonMarkovianne2}), it can be proven that $A(t) = e^{i \omega_c t} X(t)$, $B(t) = e^{i \omega_c t} Y(t)$ and $C(t) = e^{i \omega_c t} Z(t)$ are the solutions to Eq.~(\ref{dotABCtjfNonMarkovian}). Substituting $A(t)$, $B(t)$ and $C(t)$ into Eq.~(\ref{dotABCtjfNonMarkovian}), we have
\begin{equation}
\begin{aligned}
i (i \omega_c e^{i \omega_c t} X(t) + e^{i \omega_c t} \dot X(t)) = & e^{i \omega_c t} ((\omega_0-\omega_c) X(t) + g Y(t) + g Z(t)),\\
i (i \omega_c e^{i \omega_c t} Y(t) + e^{i \omega_c t} \dot Y(t)) = & e^{i \omega_c t} (g X(t) + J Z(t)) - i \int_0^t {d\tau } \int {d\omega} {\left| {\gamma_k} \right|^2}{e^{ - i (\omega - \omega_c) (t - \tau )}}e^{i \omega_c \tau} Y(\tau) \\
&- i \int_0^t {d\tau } \int {d\omega} {\left| {\gamma_k} \right|^2}{e^{ - i (\omega - \omega_c) (t - \tau )}}{e^{ i k x_1}}e^{i \omega_c \tau} Z(\tau),\\
i (i \omega_c e^{i \omega_c t} Z(t) + e^{i \omega_c t} \dot Z(t)) = & e^{i \omega_c t} (g X(t) + J Y(t)) - i \int_0^t {d\tau } \int {d\omega} {\left| {\gamma_k} \right|^2}{e^{ - i (\omega - \omega_c) (t - \tau )}}{e^{-i k x_1}}e^{i \omega_c \tau} Y(\tau) \\
&- i \int_0^t {d\tau } \int {d\omega} {\left| {\gamma_k} \right|^2}{e^{ - i (\omega - \omega_c) (t - \tau )}}e^{i \omega_c \tau} Z(\tau).
\label{APPCdotABCtjfNonMarkovian}
\end{aligned}
\end{equation}
By simplifying the above equation, Eq.~(\ref{APPCdotdtNonMarkovianne2}) can be obtained, which demonstrates that $A(t) = e^{i \omega_c t} X(t)$, $B(t) = e^{i \omega_c t} Y(t)$ and $C(t) = e^{i \omega_c t} Z(t)$ are indeed valid solutions to Eq.~(\ref{dotABCtjfNonMarkovian}). Due to $|A(t)|^2 = |e^{i \omega_c t} X(t)|^2 = |X(t)|^2$, $|B(t)|^2 = |Y(t)|^2$ and $|C(t)|^2 = |Z(t)|^2$, it suggests that the modulus squared of the probability amplitudes in the rotating frame is identical to that in the non-rotating frame. Thus, the rotation transformation does not affect the probability distribution.
\end{widetext}

\end{document}